\documentclass[journal,10pt]{IEEEtran}
\usepackage{graphicx}
\usepackage{subfigure}
\usepackage{float}
\usepackage{subfloat}
\usepackage{amsmath}
\usepackage{amssymb}
\usepackage{amsthm}
\usepackage{epstopdf}
\usepackage{cases}
\usepackage{color}
\usepackage{makecell}
\usepackage{cite}
\usepackage{algorithmic}
\usepackage[ruled]{algorithm2e}

\newtheorem{Theo}{Theorem}

\usepackage[normalem]{ulem}

\makeatletter
\renewcommand\normalsize{%
	\@setfontsize\normalsize\@xpt\@xiipt
	\abovedisplayskip 6.5\p@ \@plus2\p@ \@minus5\p@
	\abovedisplayshortskip \z@ \@plus3\p@
	\belowdisplayshortskip 6.5\p@ \@plus3\p@ \@minus3\p@
	\belowdisplayskip \abovedisplayskip
	\let\@listi\@listI}
\makeatother

\ifCLASSINFOpdf
\else
\fi
\usepackage{caption}
\usepackage{mathtools}

\begin{document}

%
\title{Kalman Filter-based Sensing in Communication Systems with Clock Asynchronism}
%
%
%

\author{Xu Chen,~\IEEEmembership{Member,~IEEE,}
	Zhiyong Feng,~\IEEEmembership{Senior Member,~IEEE,}
	J. Andrew Zhang,~\IEEEmembership{Senior Member,~IEEE,}\\
	Xin Yuan,~\IEEEmembership{Member,~IEEE,}
	and Ping Zhang,~\IEEEmembership{Fellow,~IEEE}
	\thanks{This work has been submitted to the IEEE for possible publication. Copyright may be transferred without notice, after which this version may no longer be accessible.}
	\thanks{Xu Chen and Z. Feng are with School of Information and communication Engineering, Beijing University of Posts and Telecommunications, Beijing 100876, P. R. China (Email:\{chenxu96330, fengzy\}@bupt.edu.cn).}
	\thanks{J. A. Zhang is with the Global Big Data Technologies Centre, University of Technology Sydney, Sydney, NSW 2007, Australia (Email: Andrew.Zhang@uts.edu.au).}
	\thanks{X. Yuan is with the Data61, Commonwealth Scientific and Industrial Research Organization, Sydney, NSW 2122, Australia (Email: xin.yuan@data61.csiro.au).}
	\thanks{Ping Zhang is with State Key Laboratory of Networking and Switching Technology, Beijing University of Posts and Telecommunications, Beijing 100876, P. R. China (Email: pzhang@bupt.edu.cn).}
	\thanks{Corresponding author: Zhiyong Feng}
}

%
%

\markboth{}%
{Shell \MakeLowercase{\textit{et al.}}: Bare Demo of IEEEtran.cls for IEEE Journals}
%


\maketitle

\newcounter{mytempeqncnt}
\setcounter{mytempeqncnt}{\value{equation}}
\begin{abstract}
In this paper, we propose a novel Kalman Filter (KF)-based uplink (UL) joint communication and sensing (JCAS) scheme, which can significantly reduce the range and location estimation errors due to the clock asynchronism between the base station (BS) and user equipment (UE). Clock asynchronism causes time-varying time offset (TO) and carrier frequency offset (CFO), leading to major challenges in uplink sensing. Unlike existing technologies, our scheme does not require knowing the location of the UE in advance, and retains the linearity of the sensing parameter estimation problem. We first estimate the angle-of-arrivals (AoAs) of multipaths and use them to spatially filter the CSI. Then, we propose a KF-based CSI enhancer that exploits the estimation of Doppler with CFO as the prior information to significantly suppress the time-varying noise-like TO terms in spatially filtered CSIs. Subsequently, we can estimate the accurate ranges of UE and the scatterers based on the KF-enhanced CSI. Finally, we identify the UE's AoA and range estimation and locate UE, then locate the dumb scatterers using the bi-static system. Simulation results validate the proposed scheme. The localization root mean square error of the proposed method is about 20 dB lower than the benchmarking scheme. 

\end{abstract}


%
\IEEEpeerreviewmaketitle

\section{Introduction}
%
%
%
%
\subsection{Backgrounds and Motivations}
In the future 6G networks, communication and sensing are indispensable functions for facilitating autonomous machines, such as in intelligent vehicular networks, smart factories, and smart cities~\cite{Saad2020, mahmood2020six}. Nevertheless, the rapid proliferation of wireless devices will result in severe spectrum congestion problem~\cite{liu2020joint, Chen2021CDOFDM}. Joint communications and sensing (JCAS), also known as integrated sensing and communications (ISAC), is a promising technology to solve the above problem. JCAS can achieve both sensing and communications by sharing the same transceivers and spectrum, and the same transmitted signals~\cite{Feng2021JCSC, ZhangOverviewJCS, Yuan2021}. 

Referring to the perceptive mobile networks~\cite{ZhangOverviewJCS}, sensing can be realized using both uplink (UL) and downlink (DL) signals.
UL JCAS can achieve UL communication and bi-static sensing simultaneously, without requiring full-duplex operations as in the downlink JCAS~\cite{IBFDJCR}. Therefore, UL JCAS can be realized, requiring almost no changes to network infrastructure. However, the clock asynchronism between the base station (BS) and the user equipment (UE) significantly restrains the location and Doppler frequency estimation accuracy in UL JCAS. This is because the clock asynchronism leads to time-varying timing offset (TO) and carrier frequency offset (CFO), which cause sensing ambiguity and prevent the coherent processing of sensing signals. Resolving the clock asynchronism is one of the most challenging problems in UL JCAS. 

\subsection{Related Works}

The CFO in the UL communication and localization can be estimated using conventional methods, and the residual CFO after the estimation and compensation is generally small~\cite{Zhang2022ISAC}. However, it is challenging to estimate and suppress the TO, as it is absorbed as part of the channel response and cannot be directly extracted from the received signals~\cite{Kuschel2019}.

For a long time, cooperation among multiple BSs and iterative localization using multiple transmissions is a common approach to dealing with the range and location offset due to TO. In~\cite{2019YuanLocalization}, the authors proposed an expectation-maximization (EM)-based cooperative localization method. This method requires not only multiple receivers but also 20 EM iterations, which results in a high implementation complexity. 

Recently, some new techniques have been developed to address the clock asynchronism problem. 
In~\cite{Qian2018Widar2}, the authors proposed to use the Cross-Antenna Cross-Correlation (CACC) method to achieve passive human tracking with a single WiFi link by exploiting the cross-correlation between each pair of antennas.
In~\cite{Nizhitong2021}, the authors proposed a UL JCAS method for perceptive mobile networks, allowing a static UE and BS to form a bi-static system to sense the environment by utilizing the CACC method. However, the CACC method only works under the assumption that the transmitter and receiver are static, and the accurate location of UE is known in advance. Moreover, the CACC method has to solve the challenging image targets problem. In~\cite{ZhangDaqing2019,ZhangDaqing2020,Li2022CSIsensing}, the authors proposed to use the Cross-Antenna Signal Ratio (CASR) method to estimate the Doppler frequency, using the channel state information (CSI) ratio between two antennas. However, this method works only when the scatterers are static except for the target of interest in order to maintain the linearity of the sensing parameter estimation problem based on the CSI ratio. The requirement for the prior information of the location of a static UE makes BS not able to form a bi-static sensing system with a moving UE~\cite{Nizhitong2021}, which restrains the usage of bi-static sensing in the JCAS system. Therefore, removing the assumption of knowing the accurate location of a static UE is meaningful to the realization of UL JCAS system.

\subsection{Contributions}
In this paper, referring to the perceptive mobile networks, we propose a novel Kalman Filter (KF)-based UL JCAS sensing scheme to accurately estimate the ranges and locations of the UE and dumb scatterers in the presence of clock asynchronism between bi-static UE and BS. Unlike existing techniques, our method does not require to know the location of the UE, and maintains the linearity of the sensing parameter estimation problem so that the conventional sensing algorithms, such as spectrum analysis techniques, can be applied.

We first use a two-dimensional (2D) multiple signal classification (MUSIC)-based angle-of-arrival (AoA) estimation method to estimate the AoAs of multipaths and form a spatial filter to separate the incident signals with different AoAs. Then, we propose a MUSIC-based decoupled range and Doppler estimation (DRDE) method to estimate Doppler frequency plus CFO (DPO) and ranges. A KF-based CSI enhancer is introduced during the DRDE processing to suppress the noise-like time-varying TO terms. Finally, we propose a UL bi-static localization method to first locate UE, and subsequently locate the dumb scatterers accurately using the bi-static system. 

The main contributions of this paper are summarized as follows. 
\begin{itemize}
	
	\item[1.] We propose a KF-based CSI enhancer that can use the estimated DPO as the prior information for a KF to filter the CSI and suppress the time-varying noise-like phase shift due to TO. The enhancer provides refined CSI for accurate range estimation of the UE and dumb scatterers, which can then be used for estimating the accurate location of UE. Therefore, our proposed scheme does not require knowing the accurate location of UE in advance.
	
	\item[2.] We propose a MUSIC-based DRDE method to decouple the estimation of Doppler frequency and range, estimate the DPOs as inputs to a KF-based CSI enhancer, and obtain accurate range estimation from the output of the enhancer. Moreover, the Cramer-Rao bound (CRB) for the range estimation is derived. Note that the MUSIC-based DRDE method can share the same sensing processing module with the AoA estimation.
	
	\item[3.] We propose a bi-static UL JCAS localization method for locating the UE and dumb scatterers. Exploiting the fact that the line-of-sight (LoS) path from UE to BS is the shortest, the BS first estimates the location of UE to form a bi-static system, and then uses the bi-static system to locate the dumb scatterers based on the estimated AoAs and ranges of the scatterers. 
	
\end{itemize}

We provide extensive simulation results, validating the proposed KF-based UL JCAS sensing scheme. The localization root mean square error (RMSE) of the proposed KF-based UL JCAS scheme is shown to be about 20 dB lower than the benchmarking scheme. 

The remaining parts of this paper are organized as follows. 
In section \ref{sec:system-model}, we describe the system model of the UL JCAS scheme. 
Section \ref{sec:JCAS_sensing} proposes the KF-based UL JCAS sensing scheme.
Section \ref{sec:CRB_complexity} analyzes the CRB and complexity of the proposed KF-based UL JCAS scheme.
In section \ref{sec:Simulation}, the simulation results are presented. 
Section \ref{sec:conclusion} concludes this paper.

\begin{figure}[!t]
	\centering
	\includegraphics[width=0.30\textheight]{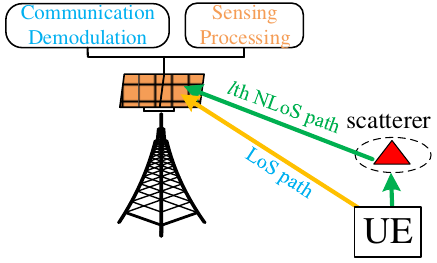}%
	\DeclareGraphicsExtensions.
	\caption{The UL JCAS scenario.}
	\label{fig: Uplink JCS Model}
\end{figure}

\textbf{Notations}: Bold uppercase letters denote matrices (e.g., $\textbf{M}$); bold lowercase letters denote column vectors (e.g., $\textbf{v}$); scalars are denoted by normal font (e.g., $\gamma$); the entries of vectors or matrices are referred to with square brackets, for instance, the $q$th entry of vector $\textbf{v}$ is $[\textbf{v}]_{q}$, and the entry of the matrix $\textbf{M}$ at the $m$th row and $q$th column is ${[\textbf{M}]_{n,m}}$; ${{\bf{U}}_s} = {\left[ {\bf{U}} \right]_{:,{N_1}:{N_2}}}$ means the matrices sliced from the $N_1$th to the $N_2$th columns of $\bf U$; $\left(\cdot\right)^H$, $\left(\cdot\right)^{*}$ and $\left(\cdot\right)^T$ denote Hermitian transpose, complex conjugate and transpose, respectively; ${\rm{vec}}(\cdot)$ is the operation to vectorize a matrix; ${\bf M}_1 \in \mathbb{C}^{M\times N}$ and ${\bf M}_2 \in \mathbb{R}^{M\times N}$ are ${M\times N}$ complex-value and real-value matrices, respectively; ${\left\| {\mathbf{v}}  \right\|_l}$ represents the ${ \ell}$-norm of ${\mathbf{v}}$, and $\ell_2$-norm is considered in this paper; for two given matrices ${\bf{S}}1$ and ${\bf{S}}2$,  ${ {[ {{v_{p,q}}} ]} |_{(p,q) \in {\bf{S}}1 \times {\bf{S}}2}}$ denotes the vector stacked by values ${v_{p,q}}$ satisfying $p\in{\bf{S}}1$ and $q\in{\bf{S}}2$; and $v \sim \mathcal{CN}(m,\sigma^2)$ means $v$ follows a circular symmetric complex Gaussian (CSCG) distribution with mean value $m$ and variance $\sigma^2$.

\begin{figure*}[!t]
	\centering
	\includegraphics[width=0.70\textheight]{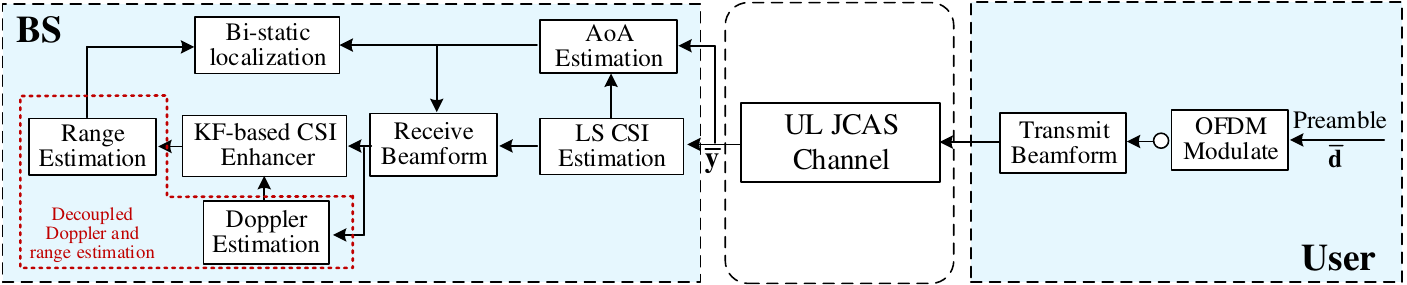}%
	\DeclareGraphicsExtensions.
	\caption{The UL JCAS scheme including the KF-based JCAS CSI estimation.}
	\label{fig: UL communication}
\end{figure*}

\section{System Model}\label{sec:system-model}
This section presents the UL JCAS system setup and channel model to provide fundamentals for UL JCAS signal processing.
\subsection{UL JCAS System Setup} \label{subsec:The Uplink JCS Model}
We consider a UL JCAS system where the BS and UE use uniform plane arrays (UPAs), as shown in Fig.~\ref{fig: Uplink JCS Model}. In the UL period, UE transmits the UL preamble and data signals, and BS uses the received training sequences in the preamble for CSI estimation and then conducts UL bi-static sensing. Using the estimated CSI, the BS also demodulates the UL data signals of UE. The data payload signals are not used for sensing in this paper. 

For simplicity of presentation, we assume that training sequences for channel estimation are transmitted at an equal interval. Each training sequence can be part of a typical packet in, e.g., WiFi systems, or transmitted in timeslot regularly, in, e.g., 5G mobile networks.  For the simplicity of notation, we refer to the packet structure in this paper, and there is one channel estimate from each packet. We also assume that within a coherent processing interval (CPI), $M_s$ CSI estimates are obtained and used for sensing, at an interval of $T_s^p$. The key idea in this paper can be extended to more general cases of non-uniform intervals. 

Orthogonal frequency division multiplexing (OFDM) signal is used. The key parameters for the OFDM signal are denoted as follows. ${P_t^U}$ is the transmit power, ${N_c}$ is the number of subcarriers occupied by UE in each OFDM symbol; $f_c$ is the carrier frequency, $T_s$ is the time duration of each symbol, and $\Delta f$ is the subcarrier interval. Since there is one CSI estimate in each packet, $T_s^p = P_s T_s$, where $P_s$ is the number of OFDM symbols in each packet.


\subsection{UPA Model} \label{subsec:UPA model}
The uniform interval between neighboring antenna elements is denoted by $d_a$. The size of UPA is ${P} \times {Q}$. The AoA for receiving or the angle-of-departure (AoD) for transmitting the $k$th far-field signal is ${{\bf{p}}_k} = {( {{\varphi _k},{\theta _k}} )^T}$, where ${\varphi _k}$ and ${\theta _k}$ are the azimuth angle and elevation angle, respectively. The phase difference between the ($p$,$q$)th antenna element and the reference antenna element is~\cite{Chen2020}
\begin{equation}\label{equ:phase_difference}
	{a_{p,q}}\left( {{{\bf{p}}_k}} \right) \! =\! \exp [ { - j\frac{{2\pi }}{\lambda }{d_a}( {p\cos {\varphi _k}\sin {\theta _k} \!+\! q\sin {\varphi _k}\sin {\theta _k}} )} ],
\end{equation}
where $\lambda = c/f_c$ is the wavelength of the carrier, $f_c$ is the carrier frequency, and $c$ is the speed of light. The steering vector for the array is given by 
\begin{equation}\label{equ:steeringVec}
	{\bf{a}}\left( {{{\bf{p}}_k}} \right) = [ {{a_{p,q}}\left( {{{\bf{p}}_k}} \right)} ]\left| {_{p = 0,1,\cdots,P - 1;q = 0,1,\cdots,Q - 1}}\right.,
\end{equation}
where $\mathbf{a}(\mathbf{p}_k) \in \mathbb{C}^{PQ \times 1}$. The sizes of the antenna arrays of the BS and user are ${P_t} \times {Q_t}$ and ${P_r} \times {Q_r}$, respectively.
 
\subsection{UL JCAS Channel Model} \label{subsec:JCAS_channel}
In this paper, we only consider propagation paths with no more than one hop, for the sake of localization. Paths with two or more hops have significantly lower power and are ignored. We assume there exists an LoS path between UE and BS. To separate different segments of a non-LoS (NLoS) path, we use subscript 1 for variables to denote the segment between UE and a scatterer and subscript 2 for the one between the scatterer and BS. For the LoS path, we use subscript 1 only.

The UL JCAS channel response at the $n$th subcarrier of the $m$th packet is given by~\cite{Zhang2022ISAC}
\begin{equation}\label{equ:H_c^U}
	{\bf{H}}_{C,n,m}{\rm{ = }}\sum\limits_{k = 0}^{K - 1} {\left[ \begin{array}{l}
			{b_{C,k}}{e^{j2\pi {{f_{c,d,k}}} mT_s^p}}{e^{ - j2\pi n\Delta {f} {{\tau _{c,k}}} }}\\
			\times {\bf{a}}( {{\bf{p}}_{R,k}^U} ){{\bf{a}}^T}( {{\bf{p}}_{T,k}^U} )
		\end{array} \right]},
\end{equation}
where ${\bf{H}}_{C,n,m} \in \mathbb{C}^{{P_t}{Q_t} \times {P_r}{Q_r}}$, $k = 0$ is for the channel response of the LoS path, and $k \in \{1, \cdots, K-1\}$ is for the $k$th NLoS path; ${\bf{a}}( {{\bf{p}}_{R,k}^U} ) \in \mathbb{C}^{{P_t}{Q_t}} \times 1$ and ${\bf{a}}( {{\bf{p}}_{T,k}^U} ) \in \mathbb{C}^{{P_r}{Q_r}} \times 1$ are the steering vectors for UL receiving and transmission, respectively; ${\bf{p}}_{R,k}^U$ and ${\bf{p}}_{T,k}^U$ are the corresponding AoA and AoD, respectively; ${f_{c,d,0}} = \frac{{{v_{0,1}}}}{\lambda }$ and ${\tau _{c,0}} = \frac{{{r_{0,1}}}}{c}$ are the Doppler shift and time delay between UE and BS of the LoS path, respectively, with ${v_{0,1}}$ and ${r_{0,1}}$ being the corresponding radial relative velocity and the distance, respectively; ${f_{c,d,k}} = {f_{d,k,1}} + {f_{d,k,2}}$ and ${\tau _{c,k}} = {\tau _{c,k,1}} + {\tau _{c,k,2}}$ are the aggregate Doppler shift and time delay of the $k$th NLoS path, respectively; ${f_{d,k,1}} = \frac{{{v_{k,1}}}}{\lambda }$ and ${f_{d,k,2}} = \frac{{{v_{k,2}}}}{\lambda }$ are the Doppler frequencies between UE and the $k$th scatterer, and between the $k$th scatterer and the BS, respectively, with ${v_{k,1}}$ and ${v_{k,2}}$ being the corresponding radial velocities; ${\tau _{c,k,1}} = \frac{{{r_{k,1}}}}{c}$ and ${\tau _{c,k,2}} = \frac{{{r_{k,2}}}}{c}$ are the time delays between UE and the $k$th scatterer, and between BS and the $k$th scatterer, respectively, with ${r_{k,1}}$ and ${r_{k,2}}$ being the corresponding distances. Moreover, ${b_{C,0}} = \sqrt {\frac{{{\lambda ^2}}}{{{{(4\pi {r_{0,1}})}^2}}}}$ and ${b_{C,k}} = \sqrt {\frac{{{\lambda ^2}}}{{{{\left( {4\pi } \right)}^3}{r_{k,1}}^2{r_{k,2}}^2}}} {\beta _{C,k}}$ are the attenuations of the LoS and NLoS paths, respectively, where ${\beta _{C,k}}$ is the reflecting factor of the $k$th scatterer, following $\mathcal{CN}(0,\sigma _{C\beta ,k}^2)$~\cite{Rodger2014principles}.

\section{UL JCAS Sensing}\label{sec:JCAS_sensing}

In this section, We present the UL JCAS sensing scheme in the presence of both TO and CFO, for which a block diagram is shown in Fig.~\ref{fig: UL communication}. We first present the initial CSI estimation using, e.g., the least-square (LS) channel estimation method. Constructing a signal correlation matrix, we show that AoA can be estimated without being impacted by TO and CFO, using an expanded 2D MUSIC-based AoA estimation method. Using the estimated AoAs, we then construct a spatial filter to separate CSIs with different AoAs. We then introduce the first two steps of the DRDE method, including decoupling the estimation of Doppler frequency and range, and estimating the DPO. We further present the KF-based CSI enhancer that uses the estimated DPOs as the prior information to suppress the time-varying noise-like TO terms in CSIs. The output from the KF-based CSI enhancer is then used to estimate the accurate ranges in the last step of DRDE. The range of the UE is identified as the smallest one among all the range estimates. Finally, BS can estimate the accurate location of the UE to form a bi-static system, based on which we propose the UL bi-static localization method to locate the scatterers.

\subsection{CSI Estimation}

The CSI may be estimated based on, e.g., the LS method using the training sequences in the preamble~\cite{2010MIMO}. We assume that a transmit beamforming (BF) with the BF vector ${\bf{w}}_{T}$ is used. Ignoring the details of training sequences and estimation method, we present the UL CSI estimated by LS method at the $n$th subcarrier of the $m$th packet as
\begin{equation}\label{equ:h_c_U_bar}
	\begin{array}{l}
		{{{\bf{\hat h}}}_{C,n,m}} = {{\bf{h}}_{C,n,m}} + {{{\bf{\bar n}}}_{t,n,m}} \in {^{{P_t}{Q_t} \times 1}}\\
		= \sqrt {P_t^U} {{\bf{H}}_{C,n,m}}{{\bf{w}}_{T}}{\zeta _{f,m}}{\zeta _{\tau,m}} + {{{\bf{\bar n}}}_{t,n,m}}\\
		= \sqrt {P_t^U} \sum\limits_{k = 0}^{K - 1} {\left[ \begin{array}{l}
				{b_{C,k}}{e^{j2\pi m{{T_s^{p}}}[{f_{c,d,k}} + {\delta _f}\left( m \right)]}}\\
				\times {e^{ - j2\pi n\Delta f[{\tau _{c,k}} + {\delta _\tau }\left( m \right)]}}\\
				\times {\chi _{T,k}}{\bf{a}}( {{\bf{p}}_{R,k}^U} )
			\end{array} \right]}  + {{{\bf{\bar n}}}_{t,n,m}},
	\end{array}
\end{equation}
where ${\bf{h}}_{C,n,m} = \sqrt {P_t^U} {\bf{H}}_{C,n,m}{\bf{w}}_{T}{\zeta _{f,m}}{\zeta _{\tau,m}}$ is the equivalent channel response, ${\zeta _{f,m}} = {e^{j2\pi m{T_s^{p}}{\delta _f}\left( m \right)}}$ and ${\zeta _{\tau,m}} = {e^{ - j2\pi n\Delta f{\delta _\tau }\left( m \right)}}$ are the phase shifts due to CFO and TO, denoted by ${\delta _f}\left( m \right)$ and ${\delta _\tau }\left( m \right)$, respectively; ${\delta _f}\left( m \right)$ and ${\delta _\tau }\left( m \right)$ are random time-varying parameters, and are assumed to follow Gaussian distribution with zero mean and variances $\sigma_f^2$ and $\sigma _\tau^2$, respectively; $T_s^p$ is the time interval between two CSI estimates; ${\bf{\bar n}}_{t,n,m} \in \mathbb{C}^{{P_t}{Q_t} \times 1}$ is the combined noise that contains Gaussian noise, and each element of
${\bf{\bar n}}_{t,n,m}$ follows $\mathcal{CN}(0,\sigma _N^2)$; ${\bf{w}}_{T}$ is the transmit BF vector with ${\left\| {{{\bf{w}}_{T}}}  \right\|_2} = 1$, and $\chi _{T,k} = {{\bf{a}}^T}( {{\bf{p}}_{T,k}^U} ) {{\bf{w}}_{T}}$ is the transmit BF gain. Then, BS can estimate the AoAs of incident signals from the CSI estimation.

\begin{algorithm}[!t]
	\caption{2D two-step Newton descent minimum searching method~\cite{2023XuJCAS}}
	\label{alg:Two-step_descent}
	\KwIn{The range of $\varphi $: ${\Phi _\varphi }$; the range of $\theta $: ${\Phi _\theta }$; the number of grid points: ${N_i}$; the maximum iteration round $ind_{max}$; the MUSIC spectrum function: $f(\bf p)$.
	}
	\KwOut{Estimation results: ${\bf{\Theta }}\!\! =\!\! { {\{ {{{{\bf{\hat p}}}_k}} \}} |_{k \in \{ {0,...,{N_s} - 1} \}}}$.}
	\textbf{Initialize: } \\
	1) ${\Phi _\varphi }$ and ${\Phi _\theta }$ are both divided evenly into ${N_i} - 1$ pieces with ${N_i}$ grid points to generate grid ${\hat \Phi _\varphi }$ and ${\hat \Phi _\theta }$.\\
	2) Set a null space ${\bf{\Theta }}$.\\
	\textbf{Process: } \\
	\textbf{Step 1}: \ForEach{${{\bf{p}}_{i,j}} \in {\hat \Phi _\varphi } \times {\Phi _\theta }$}
	{
		Calculate the spatial spectrum as ${\bf{S}}$, where ${[{\bf{S}}]_{i,j}} = [f\left( {{{\bf{p}}_{i,j}}} \right)]^{-1}$.
	}
	\textbf{Step 2}: Search the maximal values of ${\bf{S}}$ to form the set ${\bar \Theta _d}$.
	
	\textbf{Step 3}: Derive the Hessian matrix and the gradient vector of $f( {\bf{p}} )$ as ${\bf{H}}_{\bf{p}}\left( {\bf{p}} \right)$ and ${\nabla _{\bf{p}}}f\left( {\bf{p}} \right)$, respectively. 
	
	\textbf{Step 4}: \ForEach{${{\bf{p}}_{i,j}} \in {\bar \Theta _d}$}{
		\textit{k}=1\;
		${{\bf{p}}^{( 0 )}} = {{\bf{p}}_{i,j}}$\;
		${{\bf{p}}^{( k )}} = {{\bf{p}}^{\left( {k - 1} \right)}} - {[ {{{\bf{H}}_{\bf{p}}}( {{{\bf{p}}^{( {k - 1} )}}} )} ]^{ - 1}}{\nabla _{\bf{p}}}f( {{{\bf{p}}^{( {k - 1} )}}} )$\;
		\While{$\| {{{\bf{p}}^{\left( k \right)}} - {{\bf{p}}^{( {k - 1} )}}} \| > \varepsilon $ and $k \le ind_{max}$}
		{
			${{\bf{p}}^{( k )}}= {{\bf{p}}^{( {k - 1} )}} - {[ {{{\bf{H}}_{\bf{p}}}( {{{\bf{p}}^{( {k - 1} )}}} )} ]^{ - 1}}{\nabla _{\bf{p}}}f( {{{\bf{p}}^{( {k - 1} )}}} )$\;
			
		}
		${{\bf{p}}^{\left( k \right)}}$ is put into output set ${\bf{\Theta }}$\;
	}
\end{algorithm}

\subsection{AoA Estimation and Spatial Filtering} \label{sec:AoA_estimation}

\subsubsection{AoA Estimation}
We can stack all $M_s \times N_c$ CSI estimates (from $N_c$ subcarriers and $M_s$ packets) to obtain the matrix ${\bf{\hat H}}_C \in {\mathbb{C}^{{P_t}{Q_t} \times N_c M_s}}$, where the $[(m-1)N_c + n]$th column of ${\bf{\hat H}}_C$ is ${{\bf{\hat h}}_{C,n,m}}$. Construct the correlation matrix as
\begin{equation}\label{equ:R_x}
{\bf{R}}_{\bf{x}}{\rm{ = }}{{[ {{\bf{\hat H}}_C{{( {{\bf{\hat H}}_C} )}^H}} ]} \mathord{/
{\vphantom {{[ {{\bf{\hat H}}_C{{( {{\bf{\hat H}}_C} )}^H}} ]} {(M_s N_c)}}} 
\kern-\nulldelimiterspace} {(M_s N_c)}}.
\end{equation} 

Note that ${{\bf{\hat H}}_C}{({{\bf{\hat H}}_C})^H} = \sum\limits_{n,m}^{{N_c},{M_s}} {{{{\bf{\hat h}}}_{C,n,m}}{{({{{\bf{\hat h}}}_{C,n,m}})}^H}}$. According to \eqref{equ:h_c_U_bar}, we can see that ${{\bf{R}}_{\bf{x}}}$ does not contain ${e^{j2\pi m{T_s^p}{\delta _f}\left( m \right)}}{e^{ - j2\pi n\Delta f {\delta _\tau }\left( m \right)}}$. Specifically, ${e^{j2\pi m{T_s^p}{\delta _f}\left( m \right)}}{e^{ - j2\pi n\Delta f {\delta _\tau }\left( m \right)}}$ is compensated in $E\{{{\bf{h}}_{C,n,m}} {({{\bf{h}}_{C,n,m}})}^{H}\}$ given $m$; for $E\{{{\bf{h}}_{C,n,m}} {{{\bf{\bar n}}}_{t,n,m}}\}$, the expectations of the multiplication of noise and ${e^{j2\pi m{T_s^p}{\delta _f}\left( m \right)}}{e^{ - j2\pi n\Delta f {\delta _\tau }\left( m \right)}}$ are 0. Therefore, the AoA estimation is not prominently affected by the CFO and TO, which will be shown in Section~\ref{sec:Simulation}. Here, we use a refined 2D MUSIC method~\cite{2023XuJCAS} to estimate the 2D AoA based on ${\bf{R}}_{\bf{x}}$.
By applying eigenvalue decomposition to ${\bf{R}}_{\bf{x}}$, we obtain
\begin{equation}\label{equ:eigenvalue deomposition of Rx}
	[ {{\bf{U}}_x,{\bf{\Sigma }}_x} ] = \text{eig}\left( {{\bf{R}}_{\bf{x}}} \right),
\end{equation}
where $\text{eig}(\cdot)$ represents the eigenvalue decomposition of a matrix, ${\bf{\Sigma }}_x$ is the real-value eigenvalue diagonal matrix in descending order, and ${\bf{U}}_x$ is the orthogonal eigen matrix. The number of identifiable AoAs is denoted by $N_A$~\footnote{$N_A \le K$ since there may be targets with similar AoAs. The targets with similar AoA will be identified with different Doppler frequencies or ranges in the following subsections.}, which can be estimated using the minimum description length (MDL) method~\cite{2017GAOMDLMUSIC}. The noise subspace of ${{\bf{R}}_{\bf{x}}}$ is ${\bf{U}}_N = {\left[ {{\bf{U}}_x} \right]_{:,N_A + 1:{P_t}{Q_t}}}$, and then we formulate the angle spectrum function as~\cite{HAARDT2014651}
\begin{equation}\label{equ:angle_spectrum_function}
	f_a( {\bf{p}} ) = {{\bf{a}}^H}( {\bf{p}} ){\bf{U}}_N{( {{\bf{U}}_N} )^H}{\bf{a}}( {\bf{p}} ),
\end{equation}
where ${\bf{a}}\left( {\bf{p}} \right)$ is given in~\eqref{equ:steeringVec}. 
The minimum points of $f_a( {\bf{p}} )$ correspond to the AoAs to be estimated. We use a 2D two-step Newton descent method~\cite{2023XuJCAS} to derive the minimum points of $f_a( {\bf{p}} )$, which is presented in \textbf{Algorithm~\ref{alg:Two-step_descent}}. 

To identify the minimum of $f_a( {\bf{p}} )$, we substitute $f( {\bf{p}} )$, ${\bf{H}}_{\bf{p}}\left( {\bf{p}} \right)$, and ${\nabla _{\bf{p}}}f\left( {\bf{p}} \right)$ in \textbf{Algorithm~\ref{alg:Two-step_descent}} with \eqref{equ:angle_spectrum_function}, Hessian matrix and the gradient vector of $f_a( {\bf{p}} )$, respectively. Note that \textbf{Algorithm~\ref{alg:Two-step_descent}} can also be used in the one-dimensional (1D) parameter estimation by treating the second parameter to be a constant value.

The AoAs of UE and scatterers are obtained with \textbf{Algorithm~\ref{alg:Two-step_descent}} as ${\bf{\hat p}}_{R,l}^U = ( {{{\hat \varphi }_l},{{\hat \theta }_l}} )$. The direction of UE (DoU) is set as ${\bf{\hat p}}_{R,0}^U = ( {{{\hat \varphi }_0},{{\hat \theta }_0}} )$, and will be identified in Section~\ref{sec:scatter_localization} by identifying the sensing result with the smallest range. Furthermore, based on the eigenvalue vector, denoted by ${{\bf{v}}_s} = {\rm{vec}}\left( {{{\bf{\Sigma }}_x}} \right)$, obtained in the MUSIC process, we can also estimate the variance of ${{{\bf{\bar n}}}_{t,n,m}}$ as $\hat \sigma _N^2$. According to \cite{HAARDT2014651}, ${{\bf{v}}_s} \in {\mathbb{C}^{{P_t}{Q_t} \times 1}}$ can be expressed as
\begin{equation}\label{equ:vs}
	{[ {{{\bf{v}}_s}} ]_i} = \left\{ \!\!\! \begin{array}{l}
		\sigma _{s,i}^2 + \sigma _N^2,i \le N_A\\
		\sigma _N^2,i > N_A
	\end{array} \right.,
\end{equation}
where $\sigma _{s,i}^2$ is the power of the $i$th incident signal. Therefore, we obtain the estimation of the noise power as
\begin{equation}\label{equ:noise_est}
	\hat \sigma _N^2 = {{\sum\limits_{i = N_A + 1}^{{P_t}{Q_t}} {{{[ {{{\bf{v}}_s}} ]}_i}} } \mathord{/
			{\vphantom {{\sum\limits_{i = N_A + 1}^{{P_t}{Q_t}} {{{\left[ {{{\bf{v}}_s}} \right]}_i}} } {( {{P_t}{Q_t} - N_A} )}}} 
			\kern-\nulldelimiterspace} {( {{P_t}{Q_t} - N_A} )}}.
\end{equation}


\subsubsection{Spatial Filtering} \label{sec:Spatial_Filter}
In order to estimate the range of the UE and scatterers, we generate and apply a baseband spatial filter for each AoA estimate, i,e., receiving the signals in ${\bf{\hat p}}_{R,l}^U = ( {{{\hat \varphi }_l},{{\hat \theta }_l}})$. The BF vector for receiving the signals from the $l$th AoA is generated with the LS method as ${{\bf{w}}_{R,l}} = \frac{{{{\left[ {{{\bf{a}}^T}\left( {{\bf{\hat p}}_{R,l}^U} \right)} \right]}^\dag }}}{{\sqrt {\left\| {{{\left[ {{{\bf{a}}^T}\left( {{\bf{\hat p}}_{R,l}^U} \right)} \right]}^\dag }} \right\|_2^2} }}$~\cite{Zhang2019JCRS}, where $[\cdot]^{\dag}$ represents the pseudo-inverse of a matrix.
Let $\Theta _x^l$ be the index set of the targets in the $l$th AoA, where the number of elements of $\Theta _x^l$ is $N_x^l$.
 Using ${{\bf{w}}_{R,l}}$ to filter ${\bf{\hat h}}_{C,n,m}$, we obtain 
\begin{equation}\label{equ:h_nm_l}
	\begin{array}{l}
		\hat h_{C,n,m}^l{\rm{ = }}{\left( {{{\bf{w}}_{R,l}}} \right)^H}{\bf{\hat h}}_{C,n,m} = h_{C,n,m}^l + \bar n_{t,n,m}^l\\
		= \sum\limits_{k \in \Theta _x^l}^{} {\left[ \begin{array}{l}
				{e^{j2\pi m{T_s^p}{{\tilde f}_{d,k,m}}}}{e^{ - j2\pi n\Delta f{{\tilde \tau }_{k,m}}}}\\
				\sqrt {P_t^U} {b_{C,k}}{\chi _{T,k}}{\chi _{R,l,k}}
			\end{array} \right]}  + \bar n_{t,n,m}^l,
	\end{array}
\end{equation} 
where $h_{C,n,m}^l$ is the actual channel response, ${{\tilde f}_{d,k,m}} = {f_{c,d,k}} + {\delta _f}\left( m \right)$, ${{\tilde \tau }_{k,m}} = {\tau _{c,k}} + {\delta _\tau }\left( m \right)$,  and ${\chi _{R,l,k}} = {( {{{\bf{w}}_{R,l}}} )^H}{\bf{a}}( {{\bf{p}}_{R,k}^U} )$ is the receive gain of the signals from the $l$th AoA. Moreover, the equivalent interference-plus-noise term is
\begin{equation}\label{equ:n_nm_l}
	\begin{array}{l}
		\bar n_{t,n,m}^l\\
		= {\left( {{{\bf{w}}_{R,l}}} \right)^H}{{{\bf{\bar n}}}_{t,n,m}} \!\! +\!\!\!\!\!\!\! \sum\limits_{k = 0,k \notin \Theta _x^l}^{K - 1}  \!\!\left[\!\! {\begin{array}{*{20}{l}}
				{{e^{j2\pi mT_s^p{{\tilde f}_{d,k,m}}}}{e^{ - j2\pi n\Delta f{{\tilde \tau }_{k,m}}}}}\\
				{ \times \sqrt {P_t^U} {b_{C,k}}{\chi _{T,k}}{\chi _{R,I,k}}}
		\end{array}} \!\!\right],
	\end{array}
\end{equation} 
where ${\chi _{R,I,k}} = {\left( {{{\bf{w}}_{R,l}}} \right)^H}{\bf{a}}( {{\bf{p}}_{R,k}^U} ) (k \notin \Theta _x^l)$ is the receive gain of interference. It is easy to see that $\left\| {{\chi _{R,I,k}}} \right\|_2^2 \ll \left\| {{\chi _{R,l,k}}} \right\|_2^2$.

After ${\bf{\hat h}}_{C,n,m}$ at $N_c$ subcarriers of ${M_s}$ packets are all filtered by the baseband beamformer, ${{\bf{w}}_{R,l}}$, we stack all the spatially filtered CSI to form ${\bf{\hat H}}_C^l$, where ${[ {{\bf{\hat H}}_C^l} ]_{n,m}} = \hat h_{C,n,m}^l$. The actual CSI corresponding to ${{\bf{\hat H}}_C^l}$ shall be ${\bf{H}}_C^l$, where ${\left[ {{\bf{H}}_C^l} \right]_{n,m}} = h_{C,n,m}^l$.

\begin{algorithm}[!t]
	\caption{KF-based CSI Enhancer}
	\label{Kalman_CSI}
	\KwIn{The observation variance $\sigma _N^2 = \hat \sigma _N^2$; The transfer factor $A = {\hat A_{s,l}}$; The observation sequence ${{\bf{\hat h}}_C} = {\left[ {{\bf{H}}_C^l} \right]_{:,q}}$.
	}
	\KwOut{Filtered sequence ${{\bf{\bar h}}_C} = {\left[ {{\bf{H}}_C^l} \right]_{:,q}}$\; 
		\quad \quad \quad \quad The variance after KF filtering $\hat \sigma _{NK}^2$.}
	\textbf{Step} 1: The dimension of ${{\bf{\hat h}}_C}$ is obtained as $M$;
	
	\textbf{Step} 2: ${[ {{{{\bf{\bar h}}}_C}} ]_0} = {[ {{{{\bf{\hat h}}}_C}} ]_0}$; 
	
	\textbf{Step} 3: ${p_{w,0}} = {{\sum\limits_{p = 0}^{P - 1} {| {{{[ {{{{\bf{\hat h}}}_C}} ]}_p}{{( A )}^{ - p}} - {{[ {{{{\bf{\hat h}}}_C}} ]}_0}} |_2^2} } \mathord{\left/
			{\vphantom {{\sum\limits_{p = 0}^{P - 1} {\left| {{{\left[ {{{{\bf{\hat h}}}_C}} \right]}_p}{{\left( A \right)}^{ - p}} - {{\left[ {{{{\bf{\hat h}}}_C}} \right]}_0}} \right|_2^2} } P}} \right.
			\kern-\nulldelimiterspace} P}$; 
	
	\textbf{Step} 4: \For{$p$ = {\rm 1} to $M - 1$} {
		$[ {{{{\bf{\hat h}}}_C}} ]_p^ -  = A{[ {{{{\bf{\bar h}}}_C}} ]_{p - 1}}$\;
		$p_{w,p}^ -  = A{p_{w,p - 1}}{A^*}$\;
		${K_p} = {\left( {p_{w,p}^ - } \right)^*}{( {p_{w,p}^ -  + \sigma _N^2} )^{ - 1}}$\;
		${[ {{{{\bf{\bar h}}}_C}} ]_p} = [ {{{{\bf{\hat h}}}_C}} ]_p^ -  + {K_p}( {{{[ {{{{\bf{\hat h}}}_C}} ]}_p} - [ {{{{\bf{\hat h}}}_C}} ]_p^ - } )$\;
		${p_{w,p}} = \left( {1 - {K_p}} \right)p_{w,p}^ - $\;
	}
	
	\textbf{Step} 5:
	\For{$p$ = $M-1$ to 1} {
		$[ {{{{\bf{\hat h}}}_C}} ]_{p - 1}^ -  = {A^{ - 1}}{[ {{{{\bf{\bar h}}}_C}} ]_p}$\;
		$p_{w,p - 1}^ -  = {A^{ - 1}}{p_{w,p}}{( {{A^{ - 1}}} )^*}$\;
		${K_p} = {( {p_{w,p - 1}^ - } )^*}{( {p_{w,p - 1}^ -  + \sigma _N^2} )^{ - 1}}$\;
		${[ {{{{\bf{\bar h}}}_C}} ]_{p - 1}} = [ {{{{\bf{\hat h}}}_C}} ]_{p - 1}^ -  + {K_p}( {{{[ {{{{\bf{\hat h}}}_C}} ]}_{p - 1}} - [ {{{{\bf{\hat h}}}_C}} ]_{p - 1}^ - } )$\;
		${p_{w,p - 1}} = \left( {1 - {K_p}} \right)p_{w,p - 1}^ - $\;
	}
	\Return ${{{{\bf{\bar h}}}_C}}$; $\hat \sigma _{NK}^2 = {p_{w,p - 1}}$.
\end{algorithm}

\subsection{DRDE Step 1: Decoupling Range and Doppler Estimation} \label{sec:DRDE_method}

We introduce a theorem to decouple the estimation of Doppler and delay (range), which also forms the basis for the DRDE method.
Since ${\bf{\hat H}}_C^l$ has steering vector-like expressions, i.e., ${e^{j2\pi m{T_s^p}{{\tilde f}_{d,k,m}}}}$ and ${e^{ - j2\pi n\Delta f{{\tilde \tau }_{k,m}}}}$, we can construct the range and Doppler steering vectors, respectively, as
\begin{equation}\label{equ:a_r_2}
	{{\bf{a}}_{{\bf{r}},k}} = [{e^{ - j2\pi n\Delta f\frac{{{{\tilde r}_{k,m}}}}{c}}}]{|_{n = 0,1,...,{N_c} - 1}} \in {\mathbb{C}^{{N_c} \times 1}},
\end{equation}
\begin{equation}\label{equ:a_f_2}
	{{\bf{a}}_{{\bf{f}},k}} = [{e^{j2\pi m{T_s^p}{{\tilde f}_{d,k,m}}}}]{|_{m = 0,1,...,{M_s} - 1}} \in {\mathbb{C}^{{M_s} \times 1}},
\end{equation}
where ${\tilde r_{k,m}} = {\tilde \tau _{k,m}} \times c$.

Then, based on \eqref{equ:h_nm_l}, ${\bf{\hat H}}_C^l$ can be rewritten as
\begin{equation}\label{equ:H_cl_mat}
	{\bf{\hat H}}_C^l = {{\bf{a}}_{{\bf{r}},k}}{S_k}{\left( {{{\bf{a}}_{{\bf{f}},k}}} \right)^T} + {\bf{W}}_{tr}^l,
\end{equation}
where ${S_k} = \sqrt {P_t^U} {b_{C,k}}{\chi _{R,k}}{\chi _{T,k}}$, and ${\left[ {{\bf{W}}_{tr}^l} \right]_{n,m}} = \bar n_{t,n,m}^l$.
We can use \textbf{Theorem~\ref{Theo:range_Doppler}} to decouple the estimation of range and Doppler.

\begin{Theo} \label{Theo:range_Doppler}
	{\rm If ${\bf{\bar H}} = {{\bf{a}}_{\bf{r}}}\left( {{r_l}} \right)S{\left[ {{{\bf{a}}_{\bf{f}}}\left( {{f_{d,l}}} \right)} \right]^T} + {\bf{W}} \in {\mathbb{C}^{{N_c} \times {M_s}}}$, where ${{\bf{a}}_{\bf{r}}}\left( r \right) = [{e^{ - j2\pi n\Delta f\frac{r}{c}}}]{|_{n = 0,1,...,{N_c} - 1}} \in {\mathbb{C}^{{N_c} \times 1}}$, ${{\bf{a}}_{\bf{f}}}\left( {{f_d}} \right) = [{e^{j2\pi m{T_s^p}{f_d}}}]{|_{m = 0,1,...,{M_s} - 1}} \in {\mathbb{C}^{{M_s} \times 1}}$, $S$ is a complex-valued factor irrelevant to ${{\bf{a}}_{\bf{r}}}$ and ${{\bf{a}}_{\bf{f}}}$, and ${\bf{W}}$ is a Gaussian noise matrix. Let the noise subspaces of ${{\bf{\bar H}}}$ and ${ {{\bf{\bar H}}}^T}$ be ${{\bf{U}}_{x,rN}}$ and ${{\bf{U}}_{x,fN}}$, respectively. Then, the minimal values of $\| {{{\bf{U}}_{x,rN}}^H{{\bf{a}}_r}\left( r \right)} \|_2^2$ and $\| {{{\bf{U}}_{x,fN}}^H{{\bf{a}}_f}( f )} \|_2^2$ are $r = {r_l}$ and $f = {f_{d,l}}$, respectively.
		
		\begin{proof}			
			The proof is provided in \textbf{Appendix~\ref{appendix:range_Doppler}}. 
		\end{proof}
	}
\end{Theo}

Based on \textbf{Theorem~\ref{Theo:range_Doppler}}, the Doppler and range estimations are decoupled. 

\subsection{DRDE Step 2: DPO Estimation}
Based on \textbf{Theorem~\ref{Theo:range_Doppler}}, we can use ${\bf{R}}_{X,f}^l = \frac{1}{{{N_c}}}{({\bf{\hat H}}_C^l)^T}{({\bf{\hat H}}_C^l)^*}$ to derive the noise subspace of ${({\bf{\hat H}}_C^l)}^{T}$ to estimate the DPO. Here, ${\bf{R}}_{X,f}^l$ can be expressed as the average of the multiplication of the row vectors of ${\bf{\hat H}}_C^l$:
	\begin{equation}\label{equ:Rxf_vec}
		{\bf{R}}_{X,f}^l = \frac{1}{{{N_c}}}\sum\limits_{n = 1}^{{N_c}} {{{\left( {{{\left[ {{\bf{\hat H}}_C^l} \right]}_{n,:}}} \right)}^T}{{\left( {{{\left[ {{\bf{\hat H}}_C^l} \right]}_{n,:}}} \right)}^*}},
	\end{equation}
where ${{{[ {{\bf{\hat H}}_C^l} ]}_{n,:}}}$ is the $n$th row of ${{\bf{\hat H}}_C^l}$. According to \textbf{Theorem~\ref{Theo:range_Doppler}}, the MUSIC-based method can generate Doppler frequency estimates approaching most DPOs of $M_s$ packets.
By applying eigenvalue decomposition to ${\bf{R}}_{X,f}^l$, we obtain 
\begin{equation}\label{equ:EVD_of_RXf}
	\left[ {{\bf{U}}^l_{x,f},{\bf{\Sigma }}^l_{x,f}} \right] = {\text{eig}}\left( {\bf{R}}_{X,f}^l \right),
\end{equation}
where ${\bf{\Sigma }}^l_{x,f}$ is the eigenvalue diagonal matrix, and ${\bf{U}}^l_{x,f}$ is the corresponding eigen matrix. The noise subspace of ${\bf{U}}^l_{x,fN} = {[ {{\bf{U}}^l_{x,f}} ]_{:,N_{x,f}^{l}:M_s}}$, where $N_{x,f}^{l}$ is the number of scatterers with different Doppler frequencies and similar AoAs, which can be obtained using the method in \textbf{Appendix \ref{appendix:N_xU}}.

According to \textbf{Theorem~\ref{Theo:range_Doppler}}, the Doppler spectrum function and Doppler spectrum are given, respectively, by
\begin{equation}\label{equ:f_f}
	f_f\left({f}\right) = {{\bf{a}}_{\bf{f}}} {\left( f \right)^H} {\bf{U}}^l_{x,fN}{\left( {{\bf{U}}^l_{x,fN}} \right)^H}{{\bf{a}}_{\bf{f}}}\left( f \right),
\end{equation}
\begin{equation}\label{equ:S_f}
	S_f( {f}) = {[ {{{\bf{a}}_{\bf{f}}}{{( f )}^H}{\bf{U}}_{x,fN}^{l}{{( {{\bf{U}}_{x,fN}^{l}} )}^H}{{\bf{a}}_{\bf{f}}}( f )} ]^{ - 1}}.
\end{equation}

According to \textbf{Theorem~\ref{Theo:range_Doppler}}, the minimal point of $f_f\left( f \right)$, i.e., the maximal point of $S_f\left( f \right)$ is the Doppler estimation results.
\textbf{Algorithm~\ref{alg:Two-step_descent}} can be used to identify the above minimal value by treating the second parameter, $\theta$, in \textbf{Algorithm~\ref{alg:Two-step_descent}} as a discarded constant. Note that $f\left( {{{\bf{p}}}} \right)$, ${\nabla _{\bf{p}}}f\left( {{{\bf{p}}}} \right)$ and ${{{\bf{H}}_{\bf{p}}}\left( {{{\bf{p}}}} \right)}$ in \textbf{Algorithm~\ref{alg:Two-step_descent}} are replaced by \eqref{equ:f_f}, $\frac{{\partial {f_f}\left( f \right)}}{{\partial f}}$, and  $\frac{{{\partial ^2}{f_f}\left( f \right)}}{{{\partial ^2}f}}$ for Doppler estimation, respectively. 

The estimated DPOs for the targets in the $l$th AoA are denoted by $\Theta _{{n_f}}^l = {\left. {\left\{ {\hat f_{d,{n_f}}^l} \right\}} \right|_{{n_f} = 0, \cdots ,N_{x,f}^l - 1}}$. 

\subsection{KF-based CSI Enhancer}\label{sec:JCAS_CSI_Enhancer}

With the spatially filtered outputs ${{\bf{\hat H}}_C^l}$ and the estimated DPO, we now introduce the KF-based CSI enhancer to suppress the TO terms.

We can rewrite \eqref{equ:h_nm_l} as
\begin{equation}\label{equ:h_nm_l_rewrite}
	\begin{array}{l}
		\hat h_{C,n,m}^l = h_{C,n,m}^l + \bar n_{t,n,m}^l\\
		=\!\!\!\! \sum\limits_{k \in \Theta _x^l}^{} \!\!{\left[ {{A_k}{e^{j2\pi (m{T_s}{{\tilde f}_{d,k,m}} - n\Delta f{\tau _{c,k}})}}{e^{ - j2\pi n\Delta f{\delta _\tau }\left( m \right)}}} \right]}  \!+\! \bar n_{t,n,m}^l,
	\end{array}
\end{equation}
where ${A_k} = \sqrt {P_t^U} {b_{C,k}}{\chi _{T,k}}{\chi _{R,k}}$, and $\Theta _x^l$ represents the index set of the targets in the $l$th AoA as shown in Section~\ref{sec:Spatial_Filter}.

Then, we model the state transfer of the actual CSI, ${\bf{H}}_C^l$, in the $m$ axis. Let ${A_{s,\bar k}} = {e^{j2\pi T_s^p{{\tilde f}_{d,\bar k}}}}\left( {\bar k \in \Theta _x^l} \right)$ be the state transfer factor, where ${\tilde f_{d,\bar k}} = E( {{{\tilde f}_{d,\bar k,m}}} )$, then we obtain 
\begin{equation}\label{equ:h_nm_l_transfer}
	{\left[ {{\bf{H}}_C^l} \right]_{n,m + 1}} = {\left[ {{\bf{H}}_C^l} \right]_{n,m}}{A_{s,\bar k}} + {\Delta _h}\left( m \right),
\end{equation}
where ${\Delta _h}(m)$ is a time-varying noise term, and is expressed as
\begin{equation}\label{equ:delta_h}
	{\Delta _h}\left( m \right) \!\!\approx\!\! \left[\!\! \begin{array}{l}
		{A_{\bar k}}{e^{j2\pi [(m + 1){T_s^p}{{\tilde f}_{d,\bar k,m}} - n\Delta f{\tau _{c,\bar k}}]}}\\
		({e^{ - j2\pi n\Delta f{\delta _\tau }\left( {m + 1} \right)}} - {e^{ - j2\pi n\Delta f{\delta _\tau }\left( m \right)}})
	\end{array} \!\!\right] \!\!+\! \bar n_{t,n,m}^{l,\bar k},
\end{equation}
where $\bar n_{t,n,m}^{l,\bar k}$ is the time-varying error term generated by the CSI terms with $k \ne \bar k$. Moreover, \eqref{equ:delta_h} is available when $\sigma_f$ is not too large. Since CFO can be reduced to a small value according to \cite{Zhang2022ISAC}, \eqref{equ:delta_h} can be satisfied.

Based on \eqref{equ:h_nm_l_transfer} and \eqref{equ:delta_h}, there is a time-varying noise term, ${\Delta _h}(m)$ in ${\bf{\hat H}}_C^l$, containing TO in the state transfer process. We introduce a KF to filter each row of ${\bf{\hat H}}_C^l$ to suppress the time-varying TO terms, exploiting the estimated DPO, ${\hat f_{d,{n_f}}^l}$, and ${\hat A_{s,l}} = e^{j2\pi {T_s^p} {\hat f_{d,{n_f}}^l}}$
as the prior information. Note that the KF can work with a tolerance of $\sigma_f$ since \eqref{equ:delta_h} is available with a tolerance of $\sigma_f$, as will be validated in Section \ref{sec:Simulation}.

However, we cannot use KF to filter each column of ${\bf{\hat H}}_C^l$ to suppress the CFO terms. This is because given a fixed $m$, ${e^{j2\pi m{{T_s^{p}}}{\delta _f}\left( m \right)}}$ does not change with the variation of $n$. In this paper, we focus on the performance of TO suppression.

Let ${{\bf{\hat h}}_C}$ and ${{\bf{\bar h}}_C}$ denote a row of ${\bf{\hat H}}_C^l$ and its KF-filtered vector, respectively. The prior estimation of ${[ {{{{\bf{\hat h}}}_C}} ]_p}$ can be expressed as
\begin{equation}\label{equ:hc_p}
	[ {{{{\bf{\hat h}}}_C}} ]_p^ -  = {[ {{{{\bf{\hat h}}}_C}} ]_{p - 1}}A, for \; p \in \{ 1, \cdots M_s - 1\},
\end{equation}
where $A = $ $\hat A_{s,l}$ for filtering. Then, ${[ {{{{\bf{\bar h}}}_C}} ]_p}$ can be further updated as~\cite{2017Kalman}
\begin{equation}\label{equ:hc_p_bar}
	{[ {{{{\bf{\bar h}}}_C}} ]_p}{\rm{ = }}[ {{{{\bf{\hat h}}}_C}} ]_p^ - {\rm{ + }}{K_p}( {{{[ {{{{\bf{\hat h}}}_C}} ]}_p} - [ {{{{\bf{\hat h}}}_C}} ]_p^ - } ),
\end{equation}
where $K_p$ is the data fusion factor. Moreover, ${K_p}$ is expressed as~\cite{2017Kalman}
\begin{equation}\label{equ:K_p}
	{K_p} = {\left( {p_{w,p}^ - } \right)^*}{(p_{w,p}^ -  + \sigma _N^2)^{ - 1}},
\end{equation}
where $\sigma _N^2$ is the power of interference plus noise, and its estimation is $\hat \sigma _N^2$. Here, $p_{w,p}^{-}$ is the variance of the prior estimation, as given by \cite{2017Kalman}
\begin{equation}\label{equ:p_wp_}
	p_{w,p}^ -  = A{p_{w,p - 1}}{A^*},
\end{equation}
where ${p_{w,p - 1}}$ is the variance of the last estimation and given by \cite{2017Kalman}
\begin{equation}\label{equ:p_wp}
	{p_{w,p}} = \left( {1 - {K_p}} \right)p_{w,p}^ -.
\end{equation}
Specifically, ${p_{w,0}}$ is the variance of the initial observation. According to \eqref{equ:h_nm_l_transfer} and \eqref{equ:delta_h}, ${p_{w,0}}$ can be estimated as
\begin{equation}\label{equ:p_w0}
	{p_{w,0}} \approx \frac{1}{P}\sum\limits_{p = 0}^{P - 1} {|{{[{{{\bf{\hat h}}}_C}]}_p}{{ A }^{ - p}} - {{[{{{\bf{\hat h}}}_C}]}_0}|_2^2}.
\end{equation}
Based on \eqref{equ:hc_p}, \eqref{equ:hc_p_bar}, \eqref{equ:K_p}, \eqref{equ:p_wp_}, \eqref{equ:p_wp}, and \eqref{equ:p_w0}, we propose the KF-based CSI enhancer as shown in \textbf{Algorithm~\ref{Kalman_CSI}}. Note that we further add an inverse version of KF in \textbf{Step} 4 to completely exploit the sensing information, where the transfer factor is updated as ${A^{ - 1}}$. The inverse KF operation in \textbf{Step} 5 is necessary since it takes several iterations for a KF in \textbf{Step} 4 to suppress the noise terms steadily. Therefore, the initially processed terms in \textbf{Step} 4 may still contain large noise. The reverse iterations of KF in \textbf{Step} 5 can further optimize these non-ideal terms. By exploiting the estimated DPOs as the prior information, \textbf{Algorithm~\ref{Kalman_CSI}} can suppress the time-varying TO, which will be shown in Section~\ref{sec:Simulation}.

\begin{figure*}[!ht]
	\normalsize
	\setcounter{equation}{32}
	\begin{equation}\label{equ:location}
		\begin{array}{l}
			{x_n} = \left\{ \begin{array}{l}
				\frac{{{b^2}c + \sqrt {{b^4}{c^2} + {b^2}\left( {{a^2} - {c^2}} \right)\left( {{b^2} + {a^2}{{\cot }^2}{{\tilde \theta }_l} + {a^2}{{\tan }^2}{{\tilde \varphi }_l} + {a^2}{{\cot }^2}{{\tilde \theta }_l}{{\tan }^2}{{\tilde \varphi }_l}} \right)} }}{{{b^2} + {a^2}{{\cot }^2}{{\tilde \theta }_l} + {a^2}{{\tan }^2}{{\tilde \varphi }_l} + {a^2}{{\cot }^2}{{\tilde \theta }_l}{{\tan }^2}{{\tilde \varphi }_l}}},\text{for} \; \cos \left( {{{\tilde \varphi }_l}} \right) \ge 0,\\
				\frac{{{b^2}c - \sqrt {{b^4}{c^2} + {b^2}\left( {{a^2} - {c^2}} \right)\left( {{b^2} + {a^2}{{\cot }^2}{{\tilde \theta }_l} + {a^2}{{\tan }^2}{{\tilde \varphi }_l} + {a^2}{{\cot }^2}{{\tilde \theta }_l}{{\tan }^2}{{\tilde \varphi }_l}} \right)} }}{{{b^2} + {a^2}{{\cot }^2}{{\tilde \theta }_l} + {a^2}{{\tan }^2}{{\tilde \varphi }_l} + {a^2}{{\cot }^2}{{\tilde \theta }_l}{{\tan }^2}{{\tilde \varphi }_l}}}, \text{for} \; \cos \left( {{{\tilde \varphi }_l}} \right) < 0,
			\end{array} \right.\\
			{y_n} = {x_n}\tan {{\tilde \varphi }_l},\\
			{z_n} = \sqrt {\left( {x_n^2 + y_n^2} \right)\left( {1 + \frac{1}{{{{\tan }^2}{{\tilde \theta }_l}}}} \right)} \cos \left( {{{\tilde \theta }_l}} \right).
		\end{array}
	\end{equation}
	\setcounter{equation}{28}
	\hrulefill 
\end{figure*}

\subsection{DRDE Step 3: Range Estimation} \label{sec:range_estimation}

For the scatterers with the $l$th AoA and $n_f$th Doppler, the state transfer factor is $\hat A_{s,{n_f}}^l = {e^{j2\pi {T_s^p}\hat f_{d,{n_f}}^l}}$ according to Section~\ref{sec:JCAS_CSI_Enhancer}. We use \textbf{Algorithm~\ref{Kalman_CSI}} to filter all $N_c$ rows of ${\bf{\hat H}}_C^l$ by replacing 
the input with $A = \hat A_{s,{n_f}}^l = {e^{j2\pi {T_s^p}\hat f_{d,{n_f}}^l}}$ and ${{\bf{\hat h}}_C} = {[ {{\bf{\hat H}}_C^l} ]_{n,:}}$, and the filtered CSI is denoted by ${\bf{\hat H}}_{C,{n_f}}^{l,(1)}$. 

According to \textbf{Theorem~\ref{Theo:range_Doppler}}, the range can then be estimated based on ${\bf{R}}_{X,r}^{l,{n_f}} = \frac{1}{{{M_s}}}{\bf{\hat H}}_{C,{n_f}}^{l,(1)}{({\bf{\hat H}}_{C,{n_f}}^{l,(1)})^H}$. The eigenvalue decomposition of ${\bf{R}}_{X,r}^{l,{n_f}}$ is
\begin{equation}\label{equ:eig_RXr}
	\left[ {{{\bf{U}}^{l,{n_f}}_{x,r}},{{\bf{\Sigma }}^{l,{n_f}}_{x,r}}} \right] = \text{eig}\left( {{{\bf{R}}^{l,{n_f}}_{X,r}}} \right),
\end{equation}
where ${\bf{\Sigma }}^{l,{n_f}}_{x,r}$ is the eigenvalue diagonal matrix, and ${\bf{U}}^{l,{n_f}}_{x,r}$ is the corresponding eigen matrix. The noise subspace of ${{\bf{R}}^{l,{n_f}}_{{{X}},r}}$ is ${\bf{U}}_{x,rN}^{l,{n_f}} = {[ {{\bf{U}}_{x,r}^{l,{n_f}}} ]_{:,N_{x,r}^{l,n_f}:N_c}}$, where $N_{x,r}^{l,n_f}$ is the number of scatterers with different ranges, which can also be determined using the method in \textbf{Appendix~\ref{appendix:N_xU}}.

Based on \textbf{Theorem~\ref{Theo:range_Doppler}}, the range spectrum function and range spectrum are given, respectively, by
\begin{equation}\label{equ:f_r}
	f_r( {r} ) = {{\bf{a}}_r}{( r )^H}{\bf{U}}_{x,rN}^{l,{n_f}}{( {{\bf{U}}_{x,rN}^{l,{n_f}}} )^H}{{\bf{a}}_r}( r ),
\end{equation}
\begin{equation}\label{equ:S_r}
	S_r( {r} ) = {[ {{{\bf{a}}_r}{{( r )}^H}{\bf{U}}_{x,rN}^{l,{n_f}}{{( {{{\bf{U}}_{x,rN}^{l,{n_f}}}} )}^H}{{\bf{a}}_r}( r )} ]^{ - 1}}.
\end{equation}
The minimal point of $f_r\left( r \right)$, i.e., the maximal point of $S_r\left( r \right)$ corresponds to the range to be estimated.
\textbf{Algorithm~\ref{alg:Two-step_descent}} can be used to identify the above minimal value. Note that $f\left( {{{\bf{p}}}} \right)$, ${\nabla _{\bf{p}}}f\left( {{{\bf{p}}}} \right)$ and ${{{\bf{H}}_{\bf{p}}}\left( {{{\bf{p}}}} \right)}$ in \textbf{Algorithm~\ref{alg:Two-step_descent}} are replaced by \eqref{equ:f_r}, $\frac{{\partial {f_r}\left( r \right)}}{{\partial r}}$, and  $\frac{{{\partial ^2}{f_r}\left( r \right)}}{{{\partial ^2}r}}$ for range estimation, respectively.

The estimated aggregate range for the targets with the $l$th AoA and $n_f$th Doppler frequency is denoted by $\Theta _{{n_f},r}^l = {\left. {\left\{ {\hat r_{{n_f},{n_r}}^l} \right\}} \right|_{{n_r} = 0, \cdots N_{x,r}^{l,{n_f}}}}$. Note that $l = 0$ is for the range of UE.

The location of a target is not related to the Doppler frequency and only related to the AoA and range, and the range estimation method can only sense the targets in the specified $l$th AoA. Therefore, the range estimates are naturally matched with the AoA estimates and do not need to match the DPO estimates.

\subsection{UL Bi-static JCAS Localization Method}\label{sec:scatter_localization}
In this subsection, we estimate the locations of the UE and scatterers based on the estimated AoAs and ranges obtained in Section~\ref{sec:AoA_estimation} and Section~\ref{sec:range_estimation}. 

After ${\hat r_{{n_f},{n_r}}^l}$ for all $l \in \{0, \cdots, N_A-1\}$, $n_f \in \{0, \cdots, N_{x,f}^{l} -1\}$, and $n_r \in \{0, \cdots,N_{x,r}^{l,n_f} -1\}$ are estimated, the AoA and range pair for an estimated target can be combined as $\Phi _{{n_f},{n_r}}^l = \{ {\bf{\hat p}}_{R,l}^U,\hat r_{{n_f},{n_r}}^l\}$. When the SNR is high enough, the total number of estimates is $K$.

To identify the AoA and range estimates for UE, we exploit the fact that the UE's range shall be the smallest among all the range estimates. Let $\{{\bf{\hat p}}_{R,0}^U, {\hat r_{{n_f},{n_0}}^0}\}$ denote the identified AoA and range of UE, where ${\hat r_{{n_f},{n_0}}^0}$ is the smallest range estimate.

In the global coordinate system centered at BS, the location of the UE is 
\begin{equation}\label{equ:loc0}
	{{\bf{\Omega }}_0} = ( {\hat r_{{n_f},{n_0}}^0\sin {{\hat \theta }_0}\cos {{\hat \varphi }_0},\hat r_{{n_f},{n_0}}^0\sin {{\hat \theta }_0}\sin {{\hat \varphi }_0},\hat r_{{n_f},{n_0}}^0\cos {{\hat \theta }_0}} ).
\end{equation}

Construct a local coordinate system centered at BS by setting the direction from BS to UE as the positive direction of the $x$-axis. 
Then, the rotation angle from the global coordinate to the local coordinate is ${{\bf{p}}_{rot}} = ({0^{\circ} },{90^{\circ} }) - {\bf{\hat p}}_{R,0}^U$, and the $l$th AoA in the local coordinate is ${\bf{\tilde p}}_{R,l}^U = {\bf{\hat p}}_{R,l}^U{\rm{ + }}{{\bf{p}}_{rot}}$. We use ${\bf{\Omega }}_n^S = \left( {{x_n},{y_n},{z_n}} \right)$ to denote the location of the $n$th ($n = 0, \cdots, K - 1$) scatterer in the local coordinate. If the AoA and aggregate range of the $n$th scatterer is ${\Phi _n} = \left\{ {\bf{\tilde p}}_{R,l}^U, \hat r_{{n_f},{n_r}}^l \right\}$, we can obtain $x_n$, $y_n$ and $z_n$ as detailed in \textbf{Appendix~\ref{appendix:location}}. The results are provided in \eqref{equ:location}, where $a = \frac{\hat r_{{n_f},{n_r}}^l}{2}$, $c = \frac{r_{{n_f},{n_0}}^0}{2}$, and ${b^2} = {a^2} - {c^2}$. 

We can finally rotate ${\bf{\Omega }}_n^S = \left( {{x_n},{y_n},{z_n}} \right)$ into the global coordinate by rotating the angle of  $-{{\bf{p}}_{rot}}$ to obtain the location of the scatterer in the global coordinate system. 

\section{Performance Analysis of the Proposed Sensing Scheme} \label{sec:CRB_complexity}
In this section, we analyze the CRB for the KF-based range estimation method and the complexity of the KF-based sensing scheme.

\subsection{CRB for KF-based Range Estimation} \label{sec:CRB}

The ranges of the UE or scatterers are derived using $\hat h_{C,n,m}^l$. Based on \eqref{equ:h_nm_l_rewrite}, the ideal CSI expression for estimating the $k$th target is 
\setcounter{equation}{33}
\begin{equation}\label{equ:h_Cnm_l}
	\begin{array}{l}
		\hat h_{C,n,m}^{l} = {B_{n,m,k}} + \bar n_{t,n,m}^l\\
		= \left[{{e^{j2\pi m{T_s^p}{{\tilde f}_{d,k,m}}}}{e^{ - j2\pi n\Delta f \frac{{{r_k}}}{c}}}{A_k}} \right] + \bar n_{t,n,m}^l,
	\end{array}
\end{equation}
where ${A_k} = \sqrt{P_t^U} {\chi _{T,k}}{\chi _{R,k}}{b_{C,k}}$. Since ${\left\| {{{\bf{w}}_{R,l}}}  \right\|_2} = 1$, $\bar n_{t,n,m}^l = {\left( {{{\bf{w}}_{R,l}}} \right)^H}{\bf{n}}_{t,n,m}^l$ follows $\mathcal{CN}(0,\sigma_N^2)$. Therefore, $\hat h_{C,n,m}^l$ follows a Gaussian distribution with mean value ${B_{n,m,k}} = \sqrt {P_t^U} {A_k}{e^{j2\pi m{T_s^p}{{\tilde f}_{d,k,m}}}}{e^{ - j2\pi n\Delta f{{\tilde \tau }_{k,m}}}}$ and variance $\sigma_N^2$. The probability density function (PDF) of $\hat h_{C,n,m}^l$ can be expressed as
\begin{equation}\label{equ:ph}
	{p_{\hat h_{C,n,m}^l}}\!\!\!\left( h \right) \!\! =\!\! \frac{1}{{\pi \sigma _N^2}}{e^{ - \frac{1}{{\sigma _N^2}}{{\left| {h - \sqrt {P_t^U} {A_k}{e^{j2\pi m{T_s^{p}}{{\tilde f}_{d,k,m}}}}{e^{ - j2\pi n\Delta f\frac{{{r_k}}}{c}}}} \right|}^2}}}.
\end{equation}

Further, since $M_s$ OFDM symbols at $N_c$ subcarriers are used, the PDF of all the used symbols is
\begin{equation}\label{equ:ph_vec}
	{p_{\bf{h}}}\left( {\bf{h}} \right) = \frac{1}{{{{\left( {\pi \sigma _N^2} \right)}^{{N_c}{M_s}}}}} {e^{ - \frac{1}{{\sigma _N^2}}\sum\limits_{n,m}^{{N_c}{M_s}} {{{\left| {{h_{n,m}} - {B_{n,m,k}}} \right|}^2}} }}.
\end{equation}

To obtain the CRB for the range estimation of $r_k$, we need to calculate $E\left\{ {\frac{{{\partial ^2}\ln {p_{\bf{h}}}\left( {\bf{h}} \right)}}{{{\partial ^2}{r_k}}}} \right\}$. As detailed in \textbf{Appendix \ref{appendix:E_derivetive2_ph}}, we obtain 
\begin{equation}\label{equ:E_derivative2_ph}
	E\left\{ {\frac{{{\partial ^2}\ln {p_{\bf{h}}}\left( {\bf{h}} \right)}}{{{\partial ^2}{r_k}}}} \right\} =  - {\gamma _{_k}}8{\pi ^2}\Delta {f^2}\frac{1}{{{c^2}}}{M_s}\sum\limits_{n = 0}^{{N_c} - 1} {{n^2}},
\end{equation}
where ${\gamma _{_k}} = \frac{{P_t^U{{\left| {{A_k}} \right|}^2}}}{{\sigma _N^2}}$ is the received SNR.

Finally, we can obtain the CRB for the range estimation as
\begin{equation}\label{equ:CRB_range}
	{C_{{r_k}}} =  - {E^{ - 1}}\left\{ {\frac{{{\partial ^2}\ln {p_{\bf{h}}}\left( {\bf{h}} \right)}}{{{\partial ^2}{r_k}}}} \right\} = \frac{{{c^2}}}{{{\gamma _{_k}}8{\pi ^2}\Delta {f^2}{M_s}\sum\limits_{n = 0}^{{N_c} - 1} {{n^2}} }}.
\end{equation}

\begin{figure*}[!t]
	\centering
	\subfigure[The RMSEs of scatterer's AoA estimation]{\includegraphics[width=0.32\textheight]
		{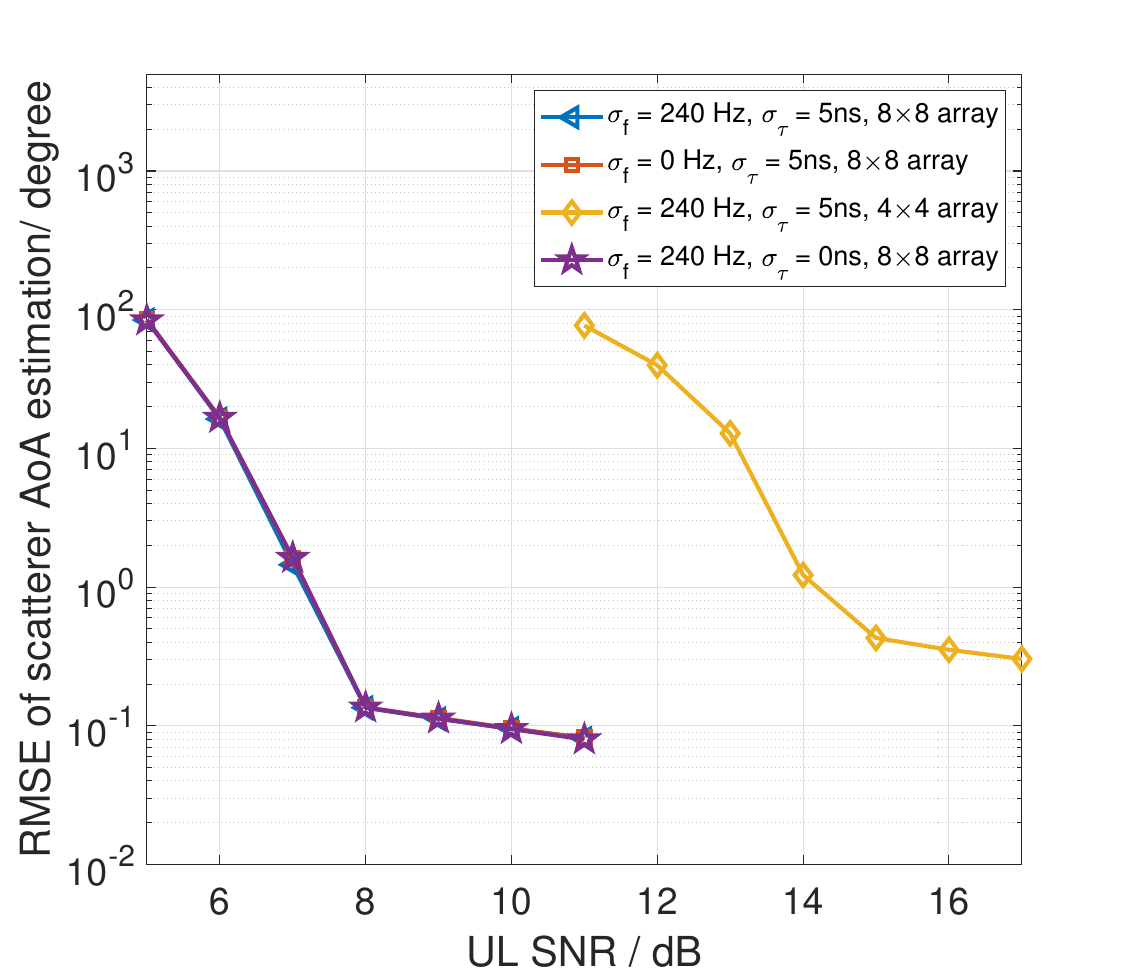}
		\label{fig: MSE_AoA_scatter}
	}
	\subfigure[The RMSEs of user's AoA estimation]{\includegraphics[width=0.32\textheight]
		{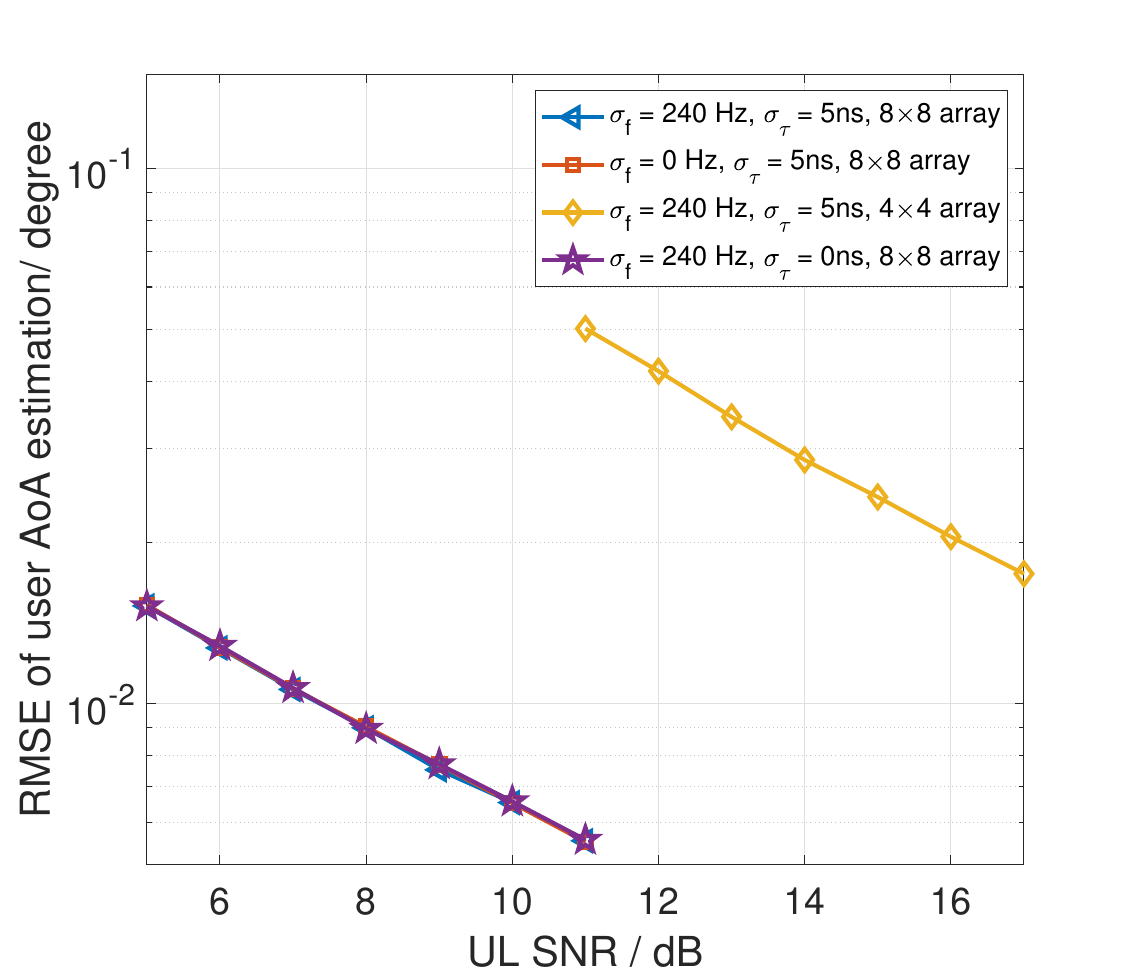}
		\label{fig: MSE_AoA_user}
	}
	\caption{The RMSEs of user and scatterer's AoAs estimation under various $\sigma_\tau$, receiver array sizes and $\sigma_f$.}
	\label{fig:MSE_AoA}
\end{figure*}

\begin{figure*}[!t]
	\centering
	\subfigure[The RMSEs of scatterer range estimation]{\includegraphics[width=0.32\textheight]
		{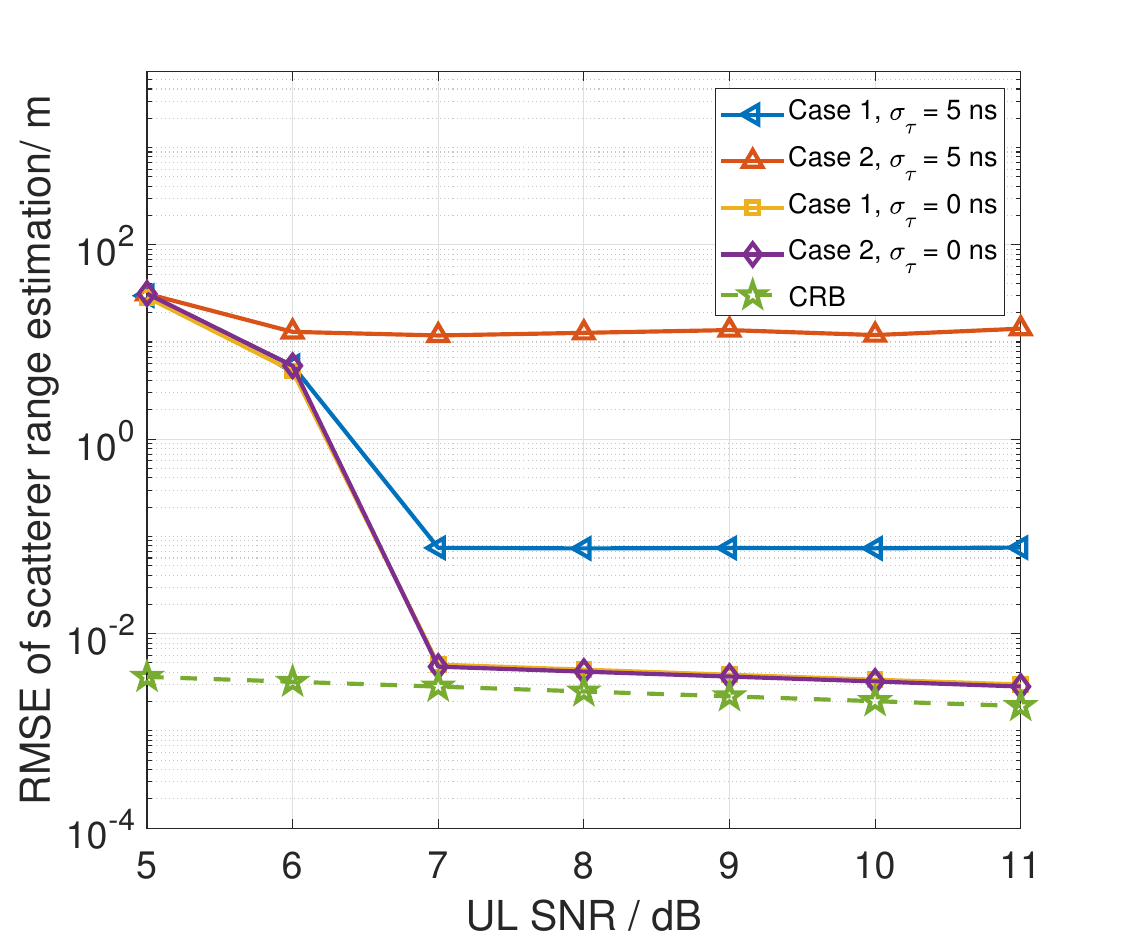}
		\label{fig: MSE_scatter_range_timing}
	}
	\subfigure[The RMSEs of scatterer localization]{\includegraphics[width=0.32\textheight]
		{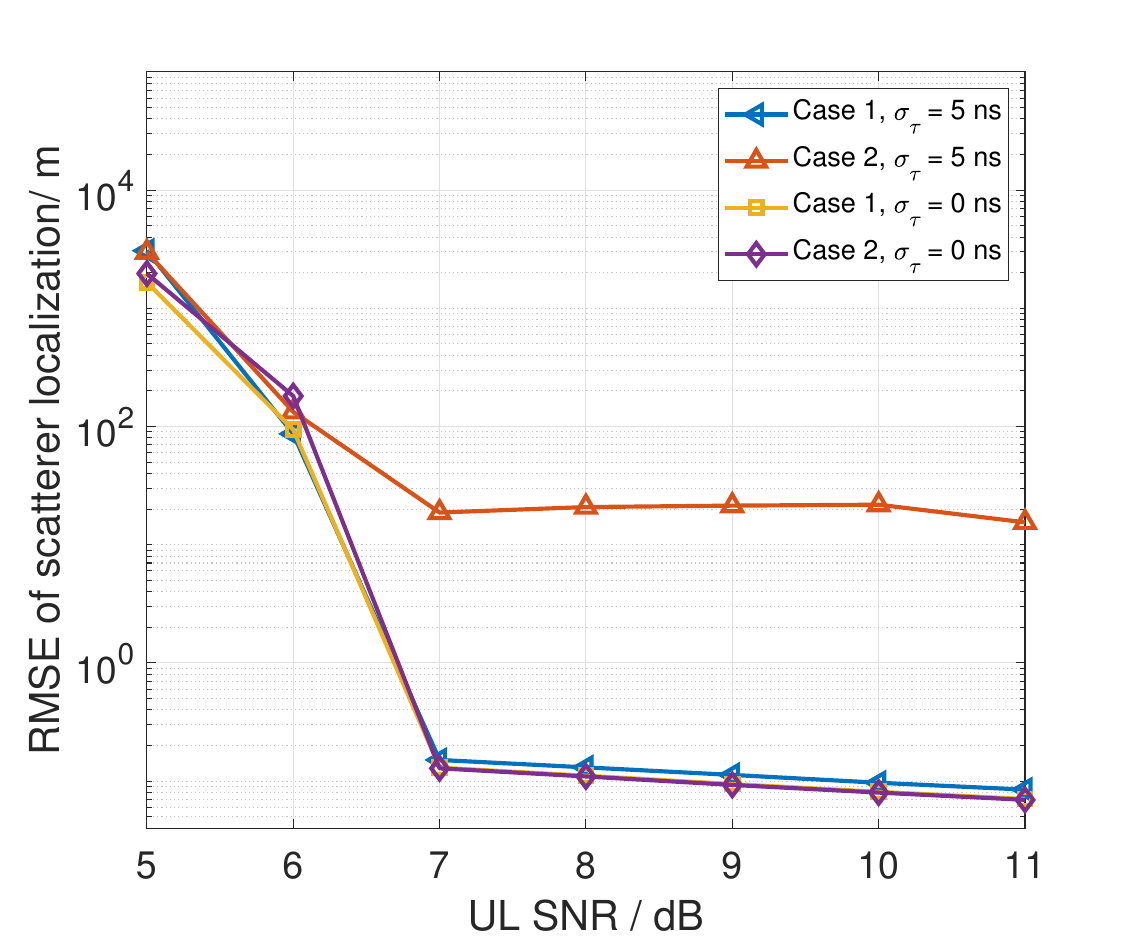}
		\label{fig: MSE_scatter_location_timing}
	}\\
	\subfigure[The RMSEs of user range estimation]{\includegraphics[width=0.32\textheight]
		{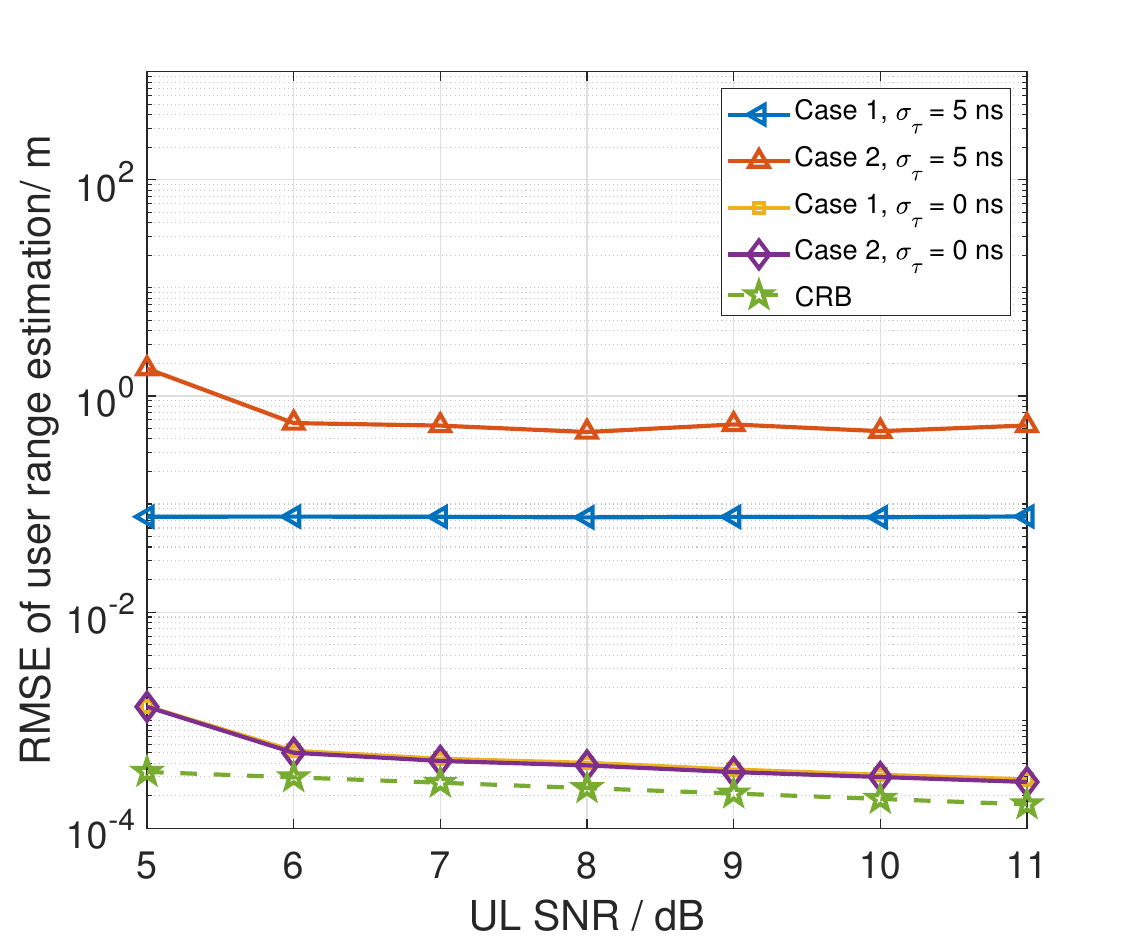}
		\label{fig: MSE_user_range_timing}
	}
	\subfigure[The RMSEs of user localization]{\includegraphics[width=0.32\textheight]
		{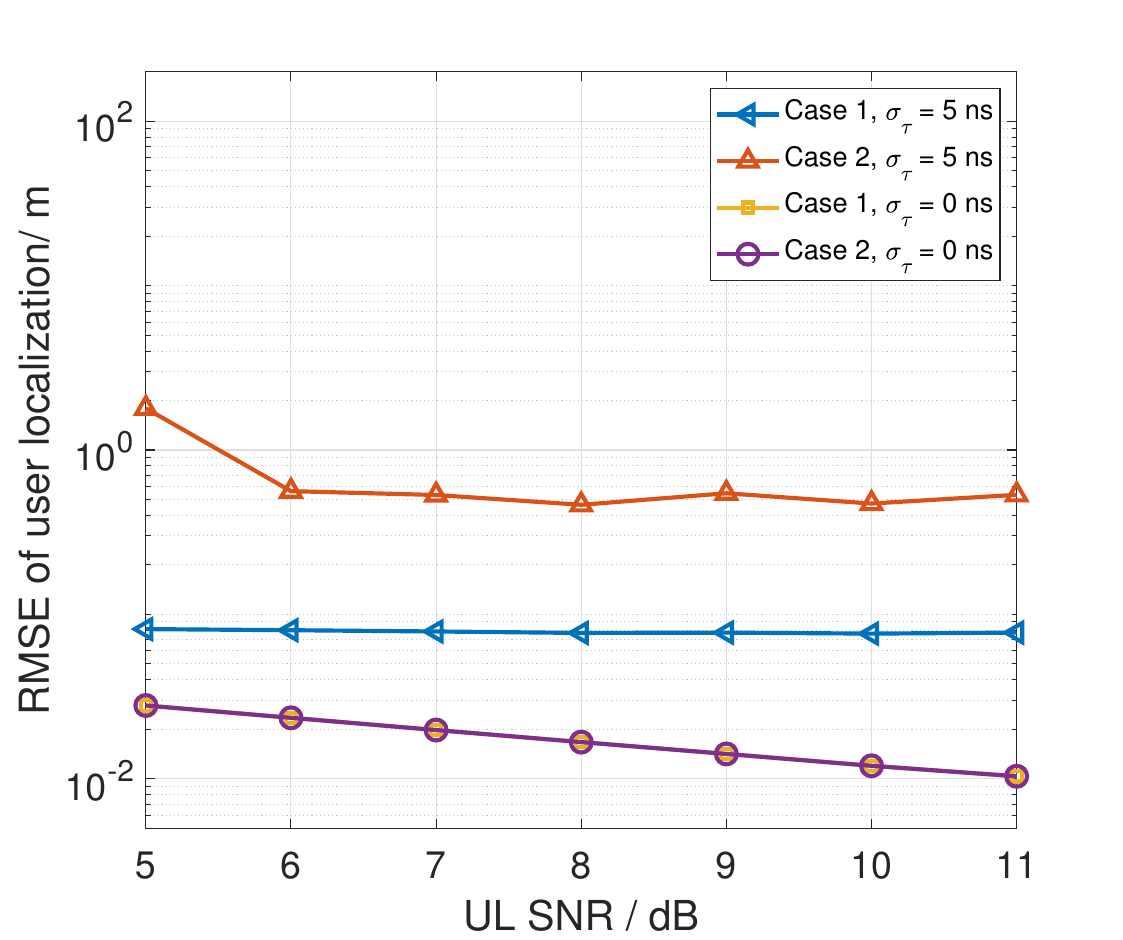}
		\label{fig: MSE_user_location_timing}
	}
	\caption{The RMSEs of range estimation and localization of \textit{cases} 1 and 2 under various $\sigma_\tau$, using $8 \times 8$ receiver array when $\sigma_f = 240$ Hz.}
	\label{fig:MSE_timing}
\end{figure*}

\subsection{Complexity of the KF-based Sensing Scheme}
The KF-based sensing scheme contains three MUSIC procedures to estimate AoA, Doppler, and range, respectively, and a KF filtering procedure for enhancing the CSI. 

The major complexity of the MUSIC procedures is from the eigenvalue decomposition and is thus $\mathcal{O}(N_{MUSIC}^3)$, where $N_{MUSIC} = \max(P_t Q_t, N_c, M_s)$, where $\max(\cdot)$ is to obtain the maximum value in a set. The KF filtering procedure only needs to add two rounds of scalar multiplications over $M_s$ CSI measurements at $N_c$ subcarriers in parallel. Therefore, the complexity of the KF filtering procedure is $\mathcal{O}(M_s)$. 

In conclusion, the complexity of the proposed KF-based sensing scheme is $\mathcal{O}(3N_{MUSIC}^3 + M_s)$, which is approximately $\mathcal{O}(N_{MUSIC}^3)$.

\section{Simulation Results}\label{sec:Simulation}
In this section, we present the simulation results for the range and location estimation RMSE of the proposed KF-based JCAS processing scheme. The global simulation parameters are listed as follows.

\subsection{System Setting}
The carrier frequency is set to 28 GHz, the antenna interval, $d_a$, is half of the wavelength, the sizes of antenna arrays of the BS and user are $P_t \times Q_t = 8 \times 8$ and $P_r \times Q_r = 1\times 1$, respectively. The subcarrier interval of UL preamble signal is $\Delta {f} =$ 480 kHz, the subcarrier number is $N_c =$ 256, and the bandwidth for JCAS is ${{B  =  }}{N_c}\Delta f = $122.88 MHz. The number of OFDM packets is set to $M_s = 64$, the number of OFDM symbols of each packet (there is only one preamble OFDM symbol for CSI estimation in each packet) is $P_s = $ 7~\cite{3GPPPhysicallayer}, and the time duration of each OFDM packet is thus $T_s^{p} = P_s/\Delta f = $ 14.58 $\mu$s.
The variance of the Gaussian noise is $\sigma_N^2 = kFTB = 4.9177\times10^{-12} $ W, where $k = 1.38 \times 10^{-23}$ J/K is the Boltzmann constant, $F = $ 10 is the noise factor, and $T = 290$ K is the standard temperature. The locations of BS and UE are (50, 4.75, 7) m and (140, 0, 2) m, respectively. The location of the dumb scatterer is (60, 3, 3) m. Moreover, we set the reflection factor of the scatterer as $\sigma _{C\beta ,k}^2 = 10$. The velocity of UE is ($-$40, 0, 0) km/h, and the velocities of BS and the dumb scatterers are (0, 0, 0) m/s. 

Note that BS does not know the locations of the UE and dumb scatterer in advance, and BS estimates the locations and ranges of the UE and dumb scatterer in each round of independent simulation. The range and location estimation MSEs are defined as the mean values of all the square errors of the range and location estimation results under a certain set of simulation parameters. The RMSE is the square root of the MSE.

Based on the above locations and velocities of BS, user, and scatterers, the AoAs, AoDs, ranges, and Doppler shifts between UE and BS, and between the scatterers and BS can be derived to generate UL channel response matrix according to the models proposed in Section~\ref{subsec:JCAS_channel}. Further, BS can use the KF-based JCAS processing scheme to estimate the ranges and locations of UE and dumb scatterers according to Section~\ref{sec:JCAS_sensing}. Communication SNR is defined as the SNR of each antenna element of BS. According to \eqref{equ:h_c_U_bar}, the UL communication SNR is expressed as
\begin{equation}\label{equ:SNR}
	{\gamma _c} = \frac{{P_t^U\sum\limits_{k = 0}^{K - 1} {{{\left| {{b_{C,k}}\chi _{T,k}} \right|}^2}} }}{{\sigma _N^2}}.
\end{equation}

\subsection{Sensing Performance}
In this subsection, we present the sensing RMSEs of the proposed KF-based JCAS sensing scheme. We predefine two cases for comparison to show the TO suppression ability of the proposed KF-based JCAS processing scheme:
	
\textit{Case 1}: The ranges of UE and scatterers are estimated by using the KF-based CSI enhancer and the DRDE method.

\textit{Case 2}: The ranges of UE and scatterers are estimated by the subspace-based super-resolution sensing method raised in \cite{2023XuJCAS}, which can use the CSI of multiple packets to estimate the range and AoA of targets and locate the targets based on the range and AoA estimation.

Figs.~\ref{fig: MSE_AoA_scatter} and \ref{fig: MSE_AoA_user} present the RMSEs of the AoA estimations of the scatterer and UE, respectively, under various $\sigma_\tau$, $\sigma_f$ and receiver array sizes. Figs.~\ref{fig: MSE_AoA_scatter} and \ref{fig: MSE_AoA_user} show that the AoA estimation RMSEs are not affected by CFO and TO, and are influenced significantly by the receive array size. As the receive array size becomes larger, the AoA estimation RMSEs decrease due to more accumulated energy for sensing. 

\begin{figure*}[!t]
	\centering
	\subfigure[The RMSEs of scatterer range estimation]{\includegraphics[width=0.32\textheight]
		{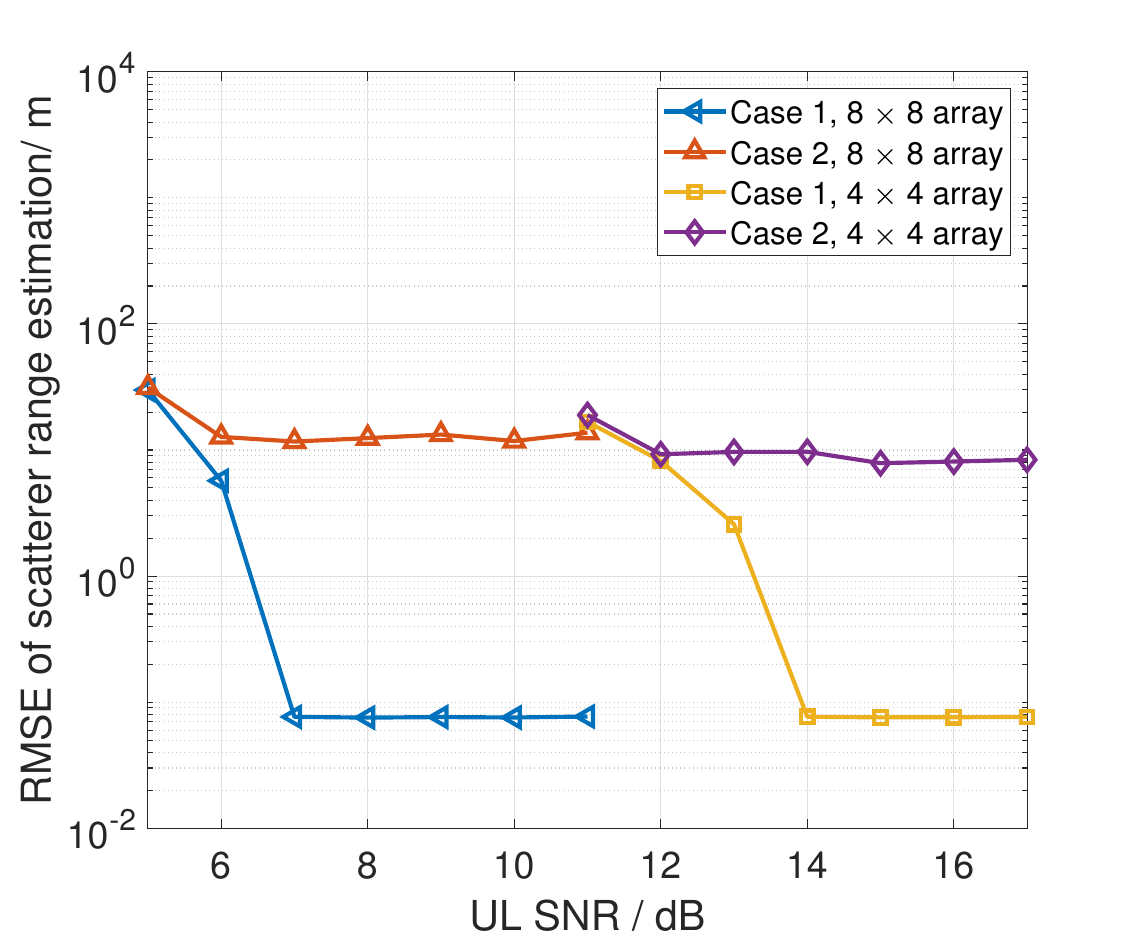}
		\label{fig: MSE_scatter_range_array}
	}
	\subfigure[The RMSEs of scatterer localization]{\includegraphics[width=0.32\textheight]
		{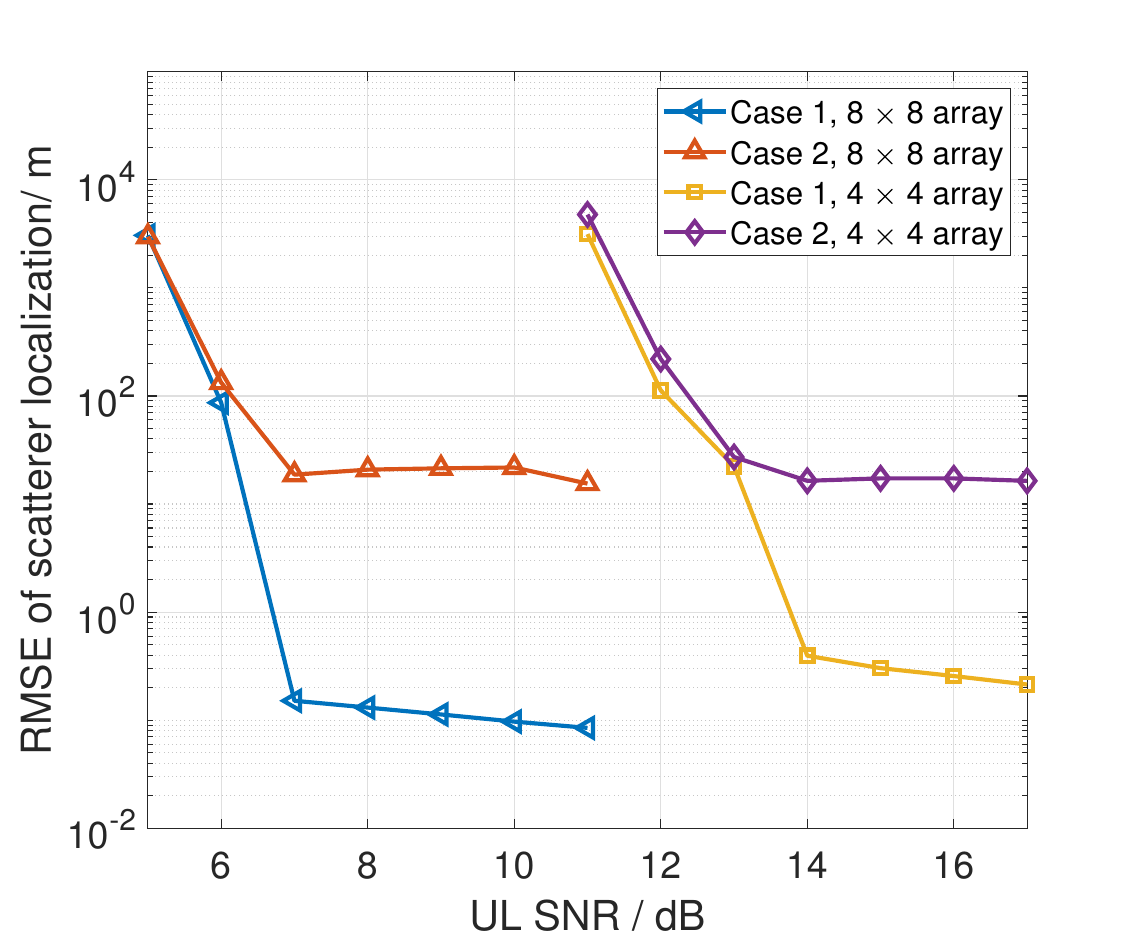}
		\label{fig: MSE_scatter_localization_array}
	}\\
	\subfigure[The RMSEs of user range estimation]{\includegraphics[width=0.32\textheight]
		{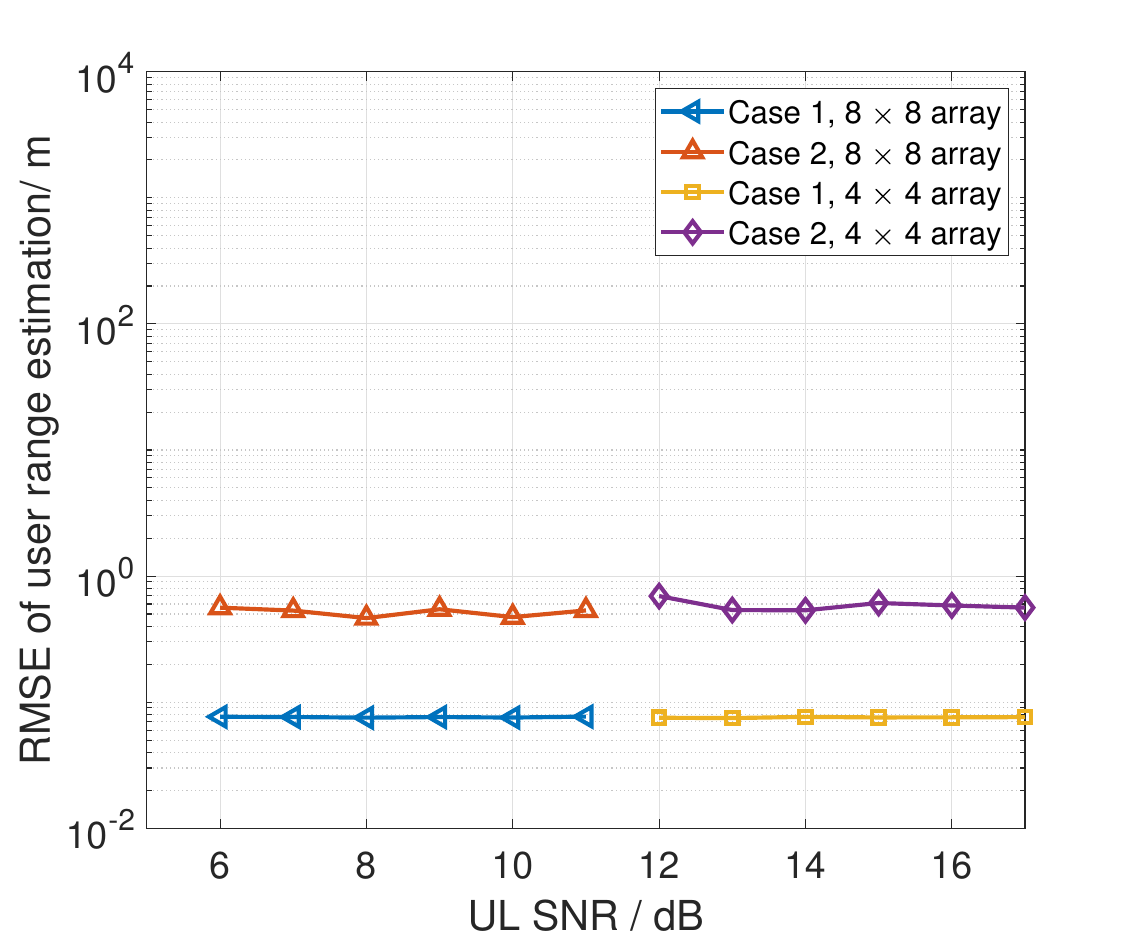}
		\label{fig: MSE_user_range_array}
	}
	\subfigure[The RMSEs of user localization]{\includegraphics[width=0.32\textheight]
		{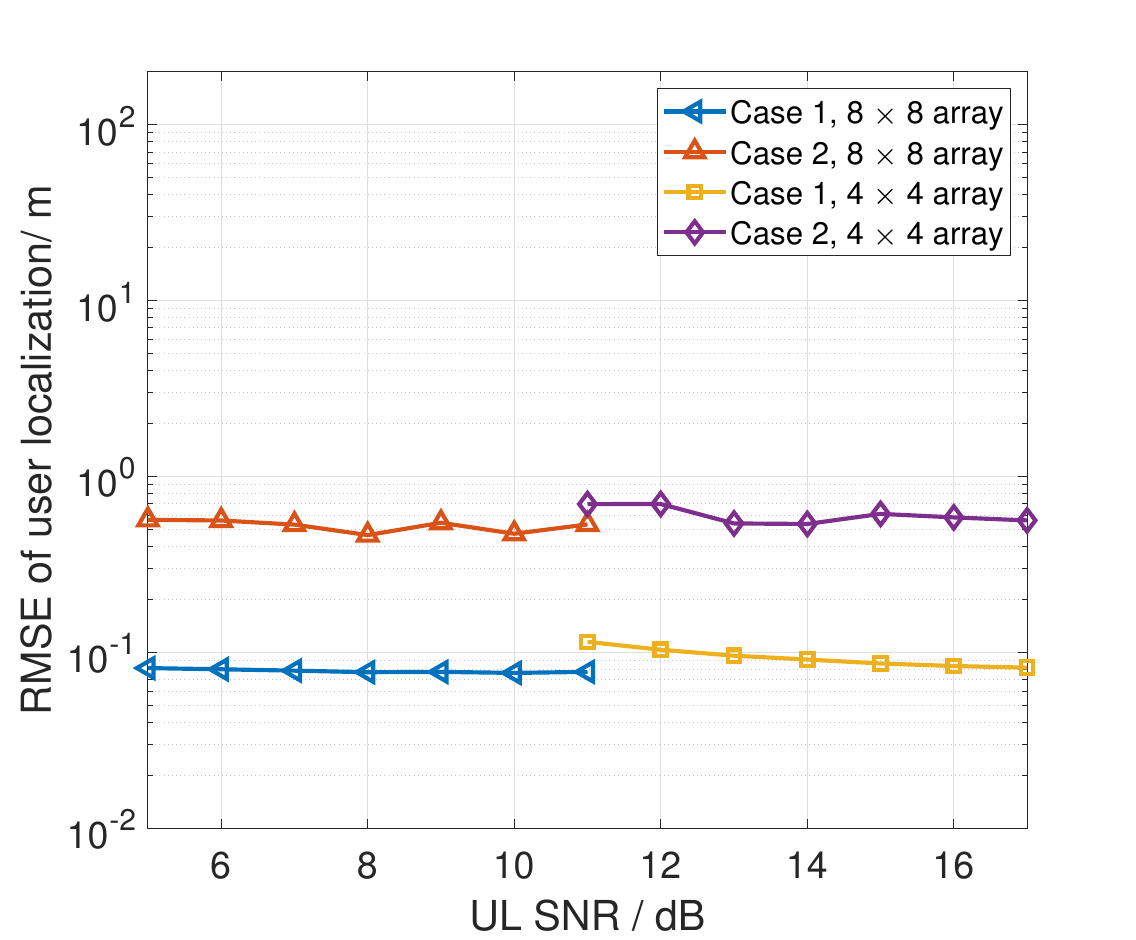}
		\label{fig: MSE_user_localization_array}
	}
	\caption{The RMSEs of range estimation and localization of \textit{cases} 1 and 2 using $4 \times 4$ and $8 \times 8$ receive arrays, when $\sigma_{\tau} = 5$ ns and $\sigma_f = 240 $ Hz.}
	\label{fig:MSE_array}
\end{figure*}

Fig.~\ref{fig:MSE_timing} shows the UE and scatterer's range and location estimation RMSEs of \textit{cases} 1 and 2, under various $\sigma_\tau$, using $8 \times 8$ receiver array when $\sigma_f = 240$ Hz. 

Figs.~\ref{fig: MSE_scatter_range_timing} and \ref{fig: MSE_user_range_timing} present the RMSEs of the scatterer and UE's range estimation RMSEs and the square roots of CRBs. Both the UE and scatterer range estimation RMSEs of \textit{case} 1 are smaller than those of \textit{case} 2 under the same SNR and $\sigma_\tau = $ 5 ns. This is because the proposed KF-based method can suppress the noise and interference better as elaborated in Section~\ref{sec:JCAS_sensing}. When $\sigma_\tau = 0$ ns, i.e., there is no TO affecting range estimation, both the UE and scatterer range estimation RMSEs of \textit{cases} 1 and 2 approach the square roots of their CRBs. When $\sigma_\tau =$ 5 ns, the proposed KF-based JCAS sensing method can still work at an acceptable RMSE level, while \textit{case} 2 cannot keep a satisfactory range estimation RMSE. From Fig.~\ref{fig: MSE_scatter_range_timing}, we can see that the RMSE results of scatterer range estimation have two stages. In the low SNR regime, the AoA estimation RMSE is large according to Fig.~\ref{fig: MSE_AoA_scatter}, which makes the spatial filter cannot separate the CSI with different AoAs well, the scatterer range estimation RMSE is thus relatively high due to the low SNR and poor spatial filter performance. When SNR is larger than 7 dB, the AoA estimation is much more accurate, and the RMSE of scatterer range estimation also becomes a small and steadily decreasing value. As for the UE range estimation shown in Fig.~\ref{fig: MSE_user_range_timing}, since the AoA estimation of UE is highly accurate in this SNR range according to Fig.~\ref{fig: MSE_AoA_user}, and the gain of LoS path is considerably larger than the NLoS path, the UE range estimation RMSEs are thus small and steady values. The gaps between the RMSEs of \textit{case} 1 and \textit{case} 2 under the same $\sigma_\tau > 0$ show the influence of the TO on the range estimation and the TO suppression ability of \textit{case} 1.

\begin{figure*}[!t]
	\centering
	\subfigure[The RMSEs of range estimation]{\includegraphics[width=0.32\textheight]
		{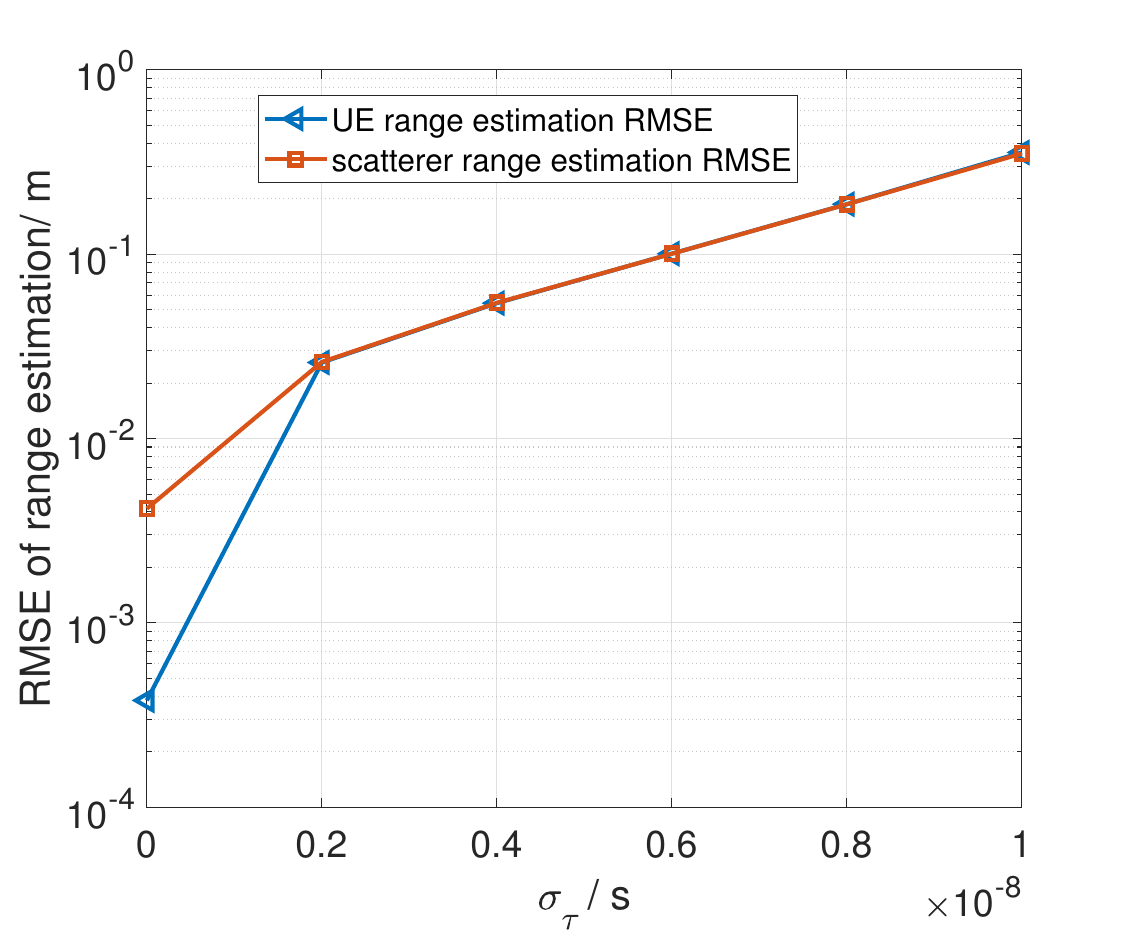}
		\label{fig: MSE_range_deltatau}
	}
	\subfigure[The RMSEs of localization]{\includegraphics[width=0.32\textheight]
		{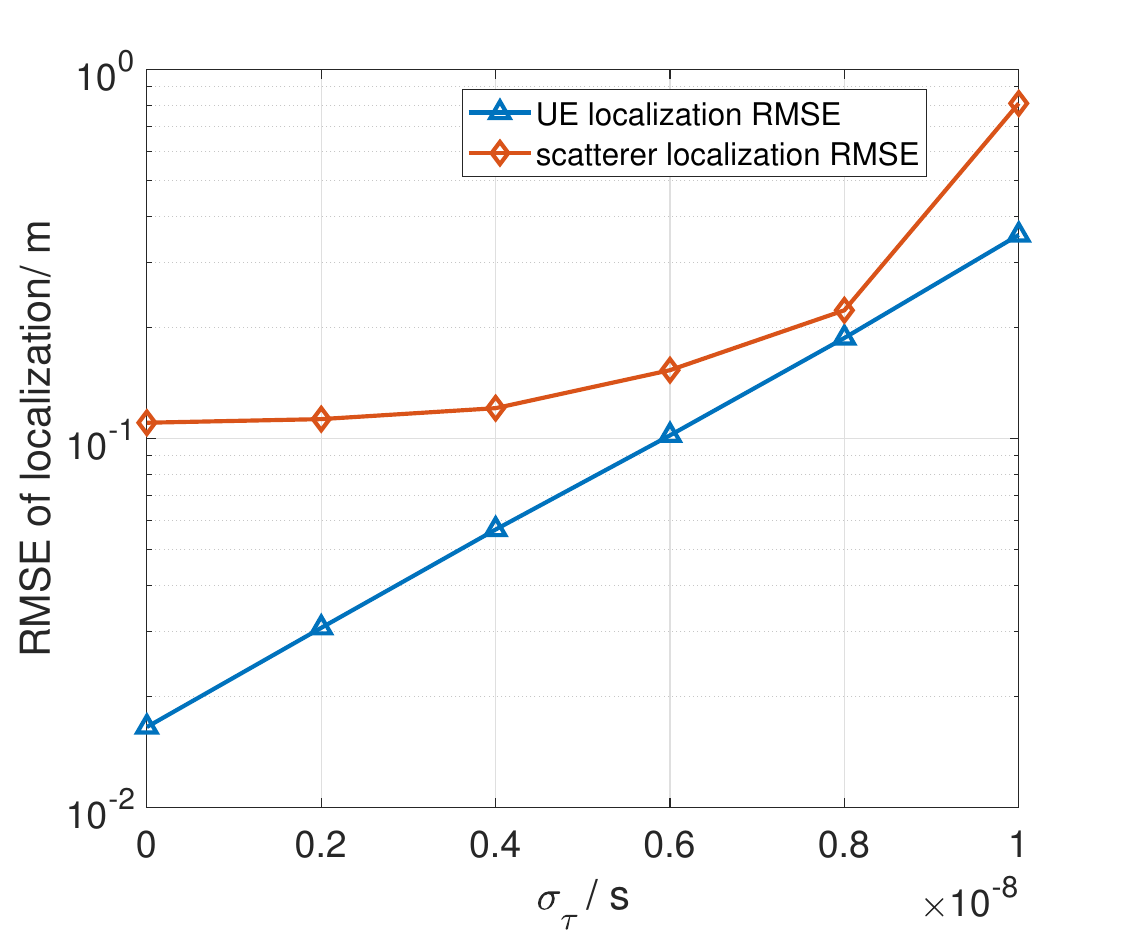}
		\label{fig: MSE_localization_deltatau}
	}
	\caption{The RMSEs of range estimation and localization of the proposed KF-based JCAS processing method changing with $\sigma_\tau$, when SNR is 8 dB, $\sigma_{f} = 100$ Hz and using $8 \times 8$ receive array.}
	\label{fig:MSE_deltatau}
\end{figure*}

Figs.~\ref{fig: MSE_scatter_location_timing} and \ref{fig: MSE_user_location_timing} present the RMSEs of the UE and scatterer's location estimation, respectively. From Fig.~\ref{fig: MSE_user_location_timing}, we can see that the UE localization RMSEs of \textit{cases} 1 and 2 are similar when $\sigma_\tau$ = 0 ns, since the UE range estimation RMSEs and AoA estimation RMSEs of two cases are approximate according to Figs. \ref{fig: MSE_user_range_timing} and \ref{fig: MSE_AoA_user}. When $\sigma_\tau$ = 5 ns, the UE localization RMSE of \textit{case} 1 is significantly smaller than that of \textit{case} 2, since the KF-based CSI enhancer of \textit{case} 1 can suppress the TO better than \textit{case} 2 as shown in Fig.~\ref{fig: MSE_user_range_timing}. As for the scatterer localization RMSEs shown in Fig.~\ref{fig: MSE_scatter_location_timing}, the scatterer localization RMSEs of \textit{case} 1 under $\sigma_\tau =$ 0 and 5 ns, and \textit{case} 2 under $\sigma_\tau =$ 0 ns, are approximate. This is because the scatterer localization accuracy is determined by the scatterer's AoA and range estimation accuracy and the UE localization accuracy. Specifically, there is an error proportional to $r \times \Delta \theta$, where $r$ is the actual range and $\Delta \theta$ is the AoA error. When the range RMSE is lower than a certain small value, the AoA error is the main factor that determines the localization RMSE. When $\sigma_\tau$ is 5 ns, \textit{case} 2 can no longer keep satisfactory scatterer and UE localization RMSEs since the UE and scatterer range estimation RMSEs of \textit{case} 2 are both considerably large. In contrast, \textit{case} 1 can still achieve acceptable localization RMSEs because the range estimation RMSEs of \textit{case} 1 are still small. The minimum scatterer localization RMSE of \textit{case} 1 is about 20 dB lower than that of \textit{case} 2 in the large SNR regime.


\begin{figure*}[!t]
	\centering
	\subfigure[The RMSEs of range estimation]{\includegraphics[width=0.32\textheight]
		{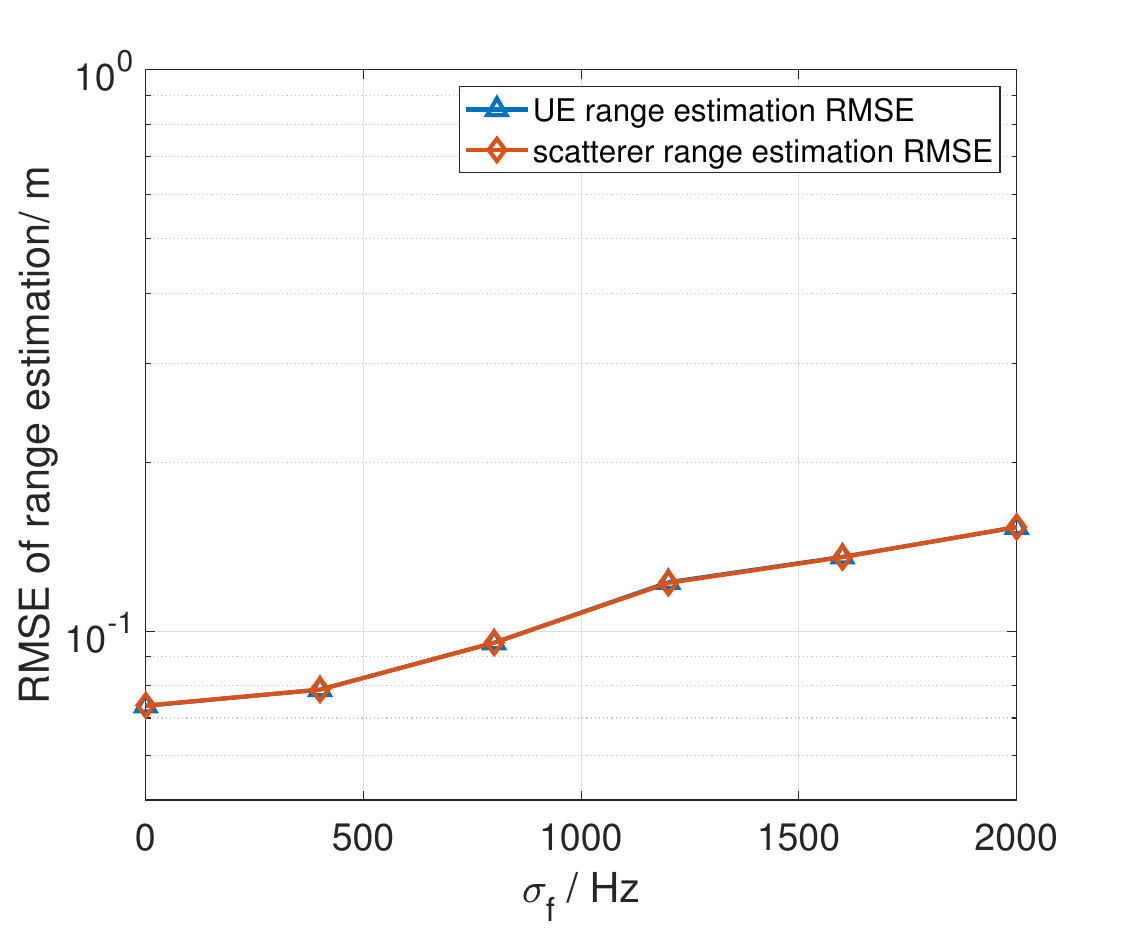}
		\label{fig: MSE_range_deltaf}
	}
	\subfigure[The RMSEs of localization]{\includegraphics[width=0.32\textheight]
		{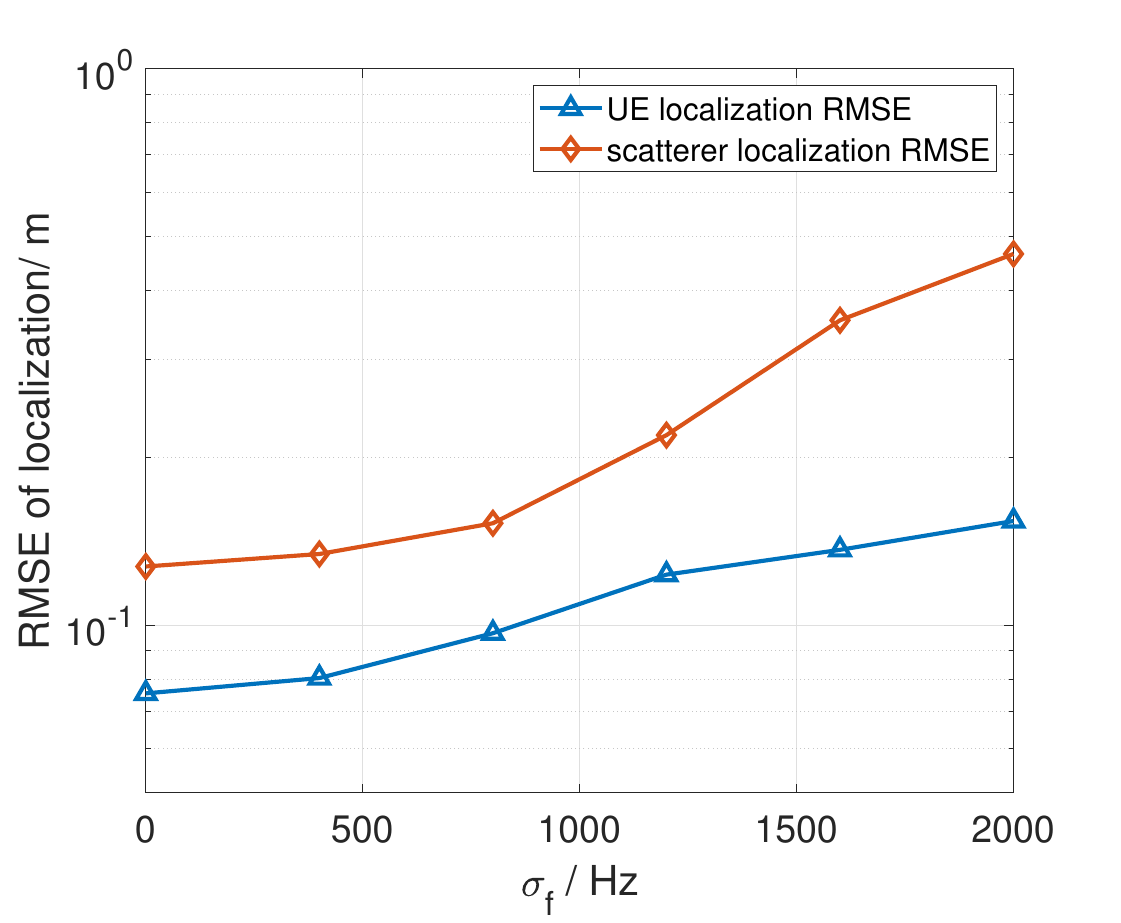}
		\label{fig: MSE_localization_deltaf}
	}
	\caption{The RMSEs of range estimation and localization of the proposed KF-based JCAS processing method changing with $\sigma_f$, when SNR is 8 dB, $\sigma_{\tau} = 5$ ns and using $8 \times 8$ receive array.}
	\label{fig:MSE_deltaf}
\end{figure*}

Fig.~\ref{fig:MSE_array} shows the UE and scatterer's range and location estimation RMSEs of \textit{cases} 1 and 2 using $4 \times 4$ and $8 \times 8$ receive arrays when $\sigma_{\tau} = 5$ ns and $\sigma_f = 240 $ Hz. 

Figs.~\ref{fig: MSE_scatter_range_array} and \ref{fig: MSE_user_range_array} plot the UE and scatterer's range estimation RMSEs. Due to the noise and TO suppression abilities of the proposed KF-based CSI enhancer, the UE and scatterer's range estimation RMSEs of \textit{case} 1 are both smaller than those of \textit{case} 2. Moreover, as the receive array size decreases from $8 \times 8$ to $4 \times 4$, the SNR required to achieve the same range estimation RMSE increases by about 6 dB. This is because the larger the array is, the more energy is accumulated for sensing. The change in array size does not influence the minimum RMSE, which shows that it only changes the processing SNR and does not have significant impact on the range estimation performance bound of the proposed method. 

Figs.~\ref{fig: MSE_scatter_localization_array} and \ref{fig: MSE_user_localization_array} present the UE and scatterer's localization RMSEs, respectively. The UE and scatterer's localization RMSEs of \textit{case} 1 are both smaller than those of \textit{case} 2 since the KF-based CSI enhancer suppresses the noise and TO. As the receive array size decreases from $8 \times 8$ to $4 \times 4$, the SNR required to achieve the same localization RMSE increases by about 7 dB. It can be seen that the SNR gap of the localization RMSEs using $8 \times 8$ and $4 \times 4$ arrays is about 1 dB larger than that of the range estimation RMSE. This is because the AoA estimation accuracy also deteriorates as the array size decreases. 

Fig.~\ref{fig:MSE_deltatau} shows the UE and scatterer's range and location estimation RMSEs of \textit{case} 1 changing with $\sigma_\tau$ from 0 ns to 10 ns, when SNR is 8 dB, $\sigma_f =$ 100 Hz, and using $8 \times 8$ receive array.

Fig.~\ref{fig: MSE_range_deltatau} plots the UE and scatterer's range estimation RMSEs, respectively. When $\sigma_\tau$ = 0 ns, the UE range estimation RMSE is smaller than the scatterer range estimation RMSE, since the gain of the LoS path is larger than that of the NLoS path. As $\sigma_\tau$ increases, the TO becomes the main factor that affects the range estimation RMSE. Therefore, the UE and scatterer range estimation RMSEs approach to a similar value restricted by the TO suppression ability of the KF-based CSI enhancer. 

Fig.~\ref{fig: MSE_localization_deltatau} plots the UE and scatterer's localization RMSEs, respectively. The UE localization RMSE increases with the increase of $\sigma_\tau$. This is because the accuracy of UE localization is affected by AoA and range estimation accuracy, and the AoA estimation RMSE is steady and small when SNR is 8 dB according to Fig.~\ref{fig:MSE_AoA}. After the range error dominates the localization error, the UE localization RMSE increases proportionally to the range estimation RMSE. In contrast, the scatterer localization accuracy is influenced by the AoA, scatterer range, and UE location estimation errors. Therefore, the scatterer localization RMSE is larger than the UE localization RMSE, and non-linearly increases with the increase of the $\sigma_\tau$.

Fig.~\ref{fig:MSE_deltaf} shows the UE and scatterer's range and location estimation RMSEs of \textit{case} 1 changing with $\sigma_f$ from 0 Hz to 2000 Hz, when SNR is 8 dB, $\sigma_\tau =$ 5 ns, and using $8 \times 8$ receive array.

Fig.~\ref{fig: MSE_range_deltaf} plots the UE and scatterer's range estimation RMSEs. The UE and scatterer range estimation RMSEs approach each other and increase slowly as $\sigma_f$ increases. This is partially due to the decoupled MUSIC-based range and Doppler estimations, which makes the change of $\sigma_f$ not have great impact on range estimation. More importantly, the proposed KF-based CSI enhancer just needs the approximate estimation of DPOs to filter the CSI, as elaborated in Section~\ref{sec:JCAS_CSI_Enhancer}. Fig.~\ref{fig: MSE_range_deltaf} shows that the proposed KF-based CSI enhancer has considerable robustness in range estimation accuracy for the inaccurate estimation of DPOs.

Fig.~\ref{fig: MSE_localization_deltaf} plots the UE and scatterer's localization RMSEs. As $\sigma_f$ increases, the UE localization RMSE increases slightly and is almost the same as the UE range estimation RMSE according to Fig.~\ref{fig: MSE_range_deltaf}. This is because the UE localization accuracy is determined by the AoA and range estimation accuracy, and the AoA estimation RMSE of UE is quite small when SNR is 8 dB according to Fig.~\ref{fig: MSE_AoA_user}. On the other hand, the scatterer localization RMSE increases more than that of the UE localization as $\sigma_f$ increases. This is because the scatterer localization RMSE is influenced by the UE localization RMSE in addition to the scatterer's range and AoA estimation RMSEs. Moreover, the scatterer AoA estimation RMSE is larger than that of the UE AoA estimation according to Fig.~\ref{fig:MSE_AoA}. Thus, the scatterer localization RMSE is larger than the UE localization RMSE.

\section{Conclusion}\label{sec:conclusion}
In this paper, we propose a KF-based UL JCAS sensing scheme that can accurately estimate the ranges and locations of UE and dumb scatterers in the presence of clock asynchronism between UE and BS. Unlike the existing solutions, our scheme works without knowing the location of UE in advance. We first estimate the AoAs of incident signals and use them to form a spatial filter to separate the CSIs with different AoAs. Then, we use the DRDE method to estimate DPOs from the spatially filtered CSIs. Subsequently, we use the KF to suppress the time-varying noise-like TO in CSIs by exploiting the estimated DPOs as the prior information. Then, the ranges of UE and scatterers are accurately estimated with the DRDE method. Moreover, the range of UE is identified as the smallest one among all the range estimates. Finally, we propose the UL bi-static JCAS localization method to accurately locate the UE and dumb scatterers by utilizing the accurately estimated ranges and AoAs. Simulation results show that the localization RMSE of the proposed KF-based UL JCAS scheme is about 20 dB lower than the method without the KF-based CSI enhancer. 

\begin{appendices} 

\section{Proof of Theorem~\ref{Theo:range_Doppler}} \label{appendix:range_Doppler}

Since ${\bf{W}}$ is Gaussian noise matrix, we assume that $E\left\{ {{\bf{W}}{{\bf{W}}^H}} \right\}{\rm{ = }}\sigma _W^2{{\bf{I}}_{{N_c}}}$ and $E\left\{ {{{\bf{W}}^H}{\bf{W}}} \right\} = \sigma _W^2{{\bf{I}}_{{M_s}}}$. The autocorrelation of ${\bf{\bar H}}$ is
\begin{equation}\label{equ:Rxr}
	\begin{array}{l}
		{{\bf{R}}_{{\bf{X}},r}} = \frac{1}{{{M_s}}}E\{ {{\bf{\bar H}}{{[ {{\bf{\bar H}}} ]}^H}} \}\\
		= S{\left( S \right)^*}{{\bf{a}}_{\bf{r}}}{\left( {{{\bf{a}}_{\bf{f}}}} \right)^T}{\left( {{{\bf{a}}_{\bf{f}}}} \right)^*}{\left( {{{\bf{a}}_{\bf{r}}}} \right)^H} + \sigma _W^2{{\bf{I}}_{{N_c}}}
	\end{array}.
\end{equation}

By applying eigenvalue decomposition to ${{\bf{R}}_{{\bf{X}},r}}$, we obtain
\begin{equation}\label{equ:Uxr}
	\left[ {{{\bf{U}}_{x,r}},{{\bf{\Sigma }}_{x,r}}} \right] = \text{eig}\left( {{{\bf{R}}_{{\bf{X}},r}}} \right),
\end{equation}
where ${{\bf{\Sigma }}_{x,r}}$ is the real-value diagonal eigenvalue matrix, ${{\bf{U}}_{x,r}}$ is the corresponding eigen matrix. Moreover, ${{\bf{U}}_{x,r}}$ can be divided as ${{\bf{U}}_{x,r}} = \left[ {{{\bf{S}}_{x,r}},{{\bf{U}}_{x,rN}}} \right]$, where  ${{{\bf{U}}_{x,rN}}}$ is the noise subspace. Because ${{\bf{U}}_{x,rN}}$ is an orthogonal unitary matrix, there are ${\left[ {{{\bf{S}}_{x,r}}} \right]^H}{{\bf{U}}_{x,rN}}{\rm{ = }}{\bf{0}}$ and ${\left[ {{{\bf{U}}_{x,rN}}} \right]^H}{{\bf{U}}_{x,rN}}{\rm{ = }}{\bf{I}}$, then we have 
\begin{equation}\label{equ:RxrUxr}
	{{\bf{R}}_{{\bf{X}},r}}{{\bf{U}}_{x,rN}} = {{\bf{U}}_{x,rN}}{{\bf{\Sigma }}_{x,r}} = \sigma _W^2{{\bf{U}}_{x,rN}}.
\end{equation}
On the other hand, according to \eqref{equ:Rxr}, we have 
\begin{equation}\label{equ:RxrUxrN2}
	\begin{aligned}
		{{\bf{R}}_{{\bf{X}},r}}{{\bf{U}}_{x,rN}} = &S{\left( S \right)^*}{{\bf{a}}_{\bf{r}}}{\left( {{{\bf{a}}_{\bf{f}}}} \right)^T}{\left( {{{\bf{a}}_{\bf{f}}}} \right)^*}{\left( {{{\bf{a}}_{\bf{r}}}} \right)^H}{{\bf{U}}_{x,rN}} \\
		&+ \sigma _W^2{{\bf{U}}_{x,rN}}.
	\end{aligned}
\end{equation}
By comparing \eqref{equ:RxrUxrN2} and \eqref{equ:RxrUxr}, we have
\begin{equation}\label{equ:conclusion_UxrN}
	S{\left( S \right)^*}{{\bf{a}}_{\bf{r}}}{\left( {{{\bf{a}}_{\bf{f}}}} \right)^T}{\left( {{{\bf{a}}_{\bf{f}}}} \right)^*}{\left( {{{\bf{a}}_{\bf{r}}}} \right)^H}{{\bf{U}}_{x,rN}} = {\bf{0}},
\end{equation}
where ${\left( {{{\bf{a}}_{\bf{f}}}} \right)^T}{\left( {{{\bf{a}}_{\bf{f}}}} \right)^*}$ is full-rank. Therefore, ${\left[ {{{\bf{U}}_{x,rN}}} \right]^H}{{\bf{a}}_{\bf{r}}}\left( {{r_l}} \right) = \;{\bf{0}}$. Thus, the minimal value of $\| {{{\bf{U}}_{x,rN}}^H{{\bf{a}}_r}\left( r \right)} \|_2^2$ is $r = {r_l}$. Similarly, we obtain the minimal value of $\| {{{\bf{U}}_{x,fN}}^H{{\bf{a}}_f}\left( f \right)} \|_2^2$ as $f = {f_{d,l}}$. This concludes the proof of \textbf{Theorem~\ref{Theo:range_Doppler}}.

\section{Derivation of $N_{x,f}^{l}$ or $N_{x,r}^{l,n_f}$} \label{appendix:N_xU}
First, we obtain the eigenvalue vector as ${{\bf{v}}_x} = {\rm{vec}}\left( {{{\bf{\Sigma }}_x}} \right)$, where ${{\bf{\Sigma }}_x} = {\bf{\Sigma }}^l_{x,f}$ or ${{\bf{\Sigma }}^{l,{n_f}}_{x,r}}$ for Doppler frequency and range estimation, respectively. Let $N$ denote the dimension of ${{\bf{v}}_x}$.
According to \cite{HAARDT2014651}, ${{\bf{v}}_x}$ can be similar to \eqref{equ:vs} with replacing $N_A$ with $N_{x,f}^{l}$ or $N_{x,r}^{l,n_f}$. We define the differential vector as ${{\bf{v}}_\Delta }$, where ${\left[ {{{\bf{v}}_\Delta }} \right]_i}{\rm{ = }}{\left[ {{{\bf{v}}_x}} \right]_i} - {\left[ {{{\bf{v}}_x}} \right]_{i + 1}}$. According to \eqref{equ:vs}, when $i>N_x$, ${\left[ {{{\bf{v}}_\Delta }} \right]_i} \approx 0$, while $i \le N_x$, ${\left[ {{{\bf{v}}_\Delta }} \right]_i} \gg 0$. Typically, we have $N_x \ll N$. Calculate the average of the latter half of ${{\bf{v}}_\Delta}$ as 
\begin{equation}\label{equ:v_ave}
	\bar v = {{\sum\limits_{k = \left\lfloor {(N - 1)/2} \right\rfloor }^{N - 1} {{{\left[ {{{\bf{v}}_\Delta }} \right]}_k}} } \mathord{\left/
			{\vphantom {{\sum\limits_{k = \left\lfloor {(N - 1)/2} \right\rfloor }^{N - 1} {{{\left[ {{{\bf{v}}_\Delta }} \right]}_k}} } {\left( {N - \left\lfloor {(N - 1)/2} \right\rfloor } \right)}}} \right.
			\kern-\nulldelimiterspace} {\left( {N - \left\lfloor {(N - 1)/2} \right\rfloor } \right)}},
\end{equation}
where $\left\lfloor \cdot \right\rfloor$ is the floor function. Here, $\bar v$ shall be an extremely small value approaching 0. Therefore, according to the maximum likelihood criterion, $N_{x,f}^{l}$ or $N_{x,r}^{l,n_f}$ can be estimated as 
\begin{equation}\label{equ:N_x_est}
	{\hat N}_x = \mathop {\arg \max }\limits_i {\left[ {{{\bf{v}}_\Delta }} \right]_i} > \left( {1 + \varepsilon } \right)\bar v,
\end{equation}
where $\varepsilon $ is a parameter used to avoid false estimation due to the small error. In this paper, we set $\varepsilon = 1$.

\section{Derivation of \eqref{equ:location}}\label{appendix:location}
In the local coordinate, construct the ellipsoid by setting BS and UE as the focuses. The location of BS in the local coordinate is $(0,0,0)$. The location of the scatterer is the intersection between the radial ray pointing at ${\bf{\tilde p}}_{R,l}^U = \left( {{{\tilde \varphi }_l},{{\tilde \theta }_l}} \right)$ and the ellipsoid. Therefore, we can construct the equation set as
\begin{equation}\label{equ:location_set}
	\left\{ \begin{array}{l}
		\frac{{{{\left( {{x_n} - c} \right)}^2}}}{{{a^2}}} + \frac{{y_n^2 + z_n^2}}{{{b^2}}} = 1\\
		{y_n} = {x_n}\tan {{\tilde \varphi }_l}\\
		{z_n} = \sqrt {\left( {x_n^2 + y_n^2 + z_n^2} \right)} \cos {{\tilde \theta }_l}
	\end{array} \right.,
\end{equation}
where $a = \frac{\hat r_{{n_f},{n_r}}^l}{2}$, $c = \frac{r_{{n_f},{n_0}}^0}{2}$, and ${b^2} = {a^2} - {c^2}$. By solving \eqref{equ:location_set}, we can obtain \eqref{equ:location}.

\section{Derivation of \eqref{equ:E_derivative2_ph}}\label{appendix:E_derivetive2_ph}

According to \eqref{equ:ph_vec}, we obtain 
\begin{equation}\label{equ:ln_ph}
	\ln {p_{\bf{h}}}\left( {\bf{h}} \right) \! = \! - {N_c}{M_s}\ln \left( {\pi \sigma _N^2} \right)\! -\! \frac{1}{{\sigma _N^2}}\sum\limits_{n,m}^{{N_c}{M_s}} {{{\left| {{h_{n,m}} - {B_{n,m,k}}} \right|}^2}}.
\end{equation}

Then, the first-order derivative of $\ln {p_{\bf{h}}}\left( {\bf{h}} \right)$ is obtained as
\begin{equation}\label{equ:lnph_firstorder}
	\frac{{\partial \ln {p_{\bf{h}}}\left( {\bf{h}} \right)}}{{\partial {r_k}}} = \frac{1}{{\sigma _N^2}}\sum\limits_{n,m}^{} {\left\{ \begin{array}{l}
			\left( {{h_{n,m}} - {B_{n,m,k}}} \right)\frac{{\partial {{\left( {{B_{n,m,k}}} \right)}^*}}}{{\partial {r_l}}}\\
			+ \frac{{\partial {B_{n,m,k}}}}{{\partial {r_l}}}\left( {h_{n,m}^* - B_{n,m,k}^*} \right)
		\end{array} \right\}}.
\end{equation}

Since we have 
\begin{equation}\label{equ:B_firstorder_feature}
	\frac{{\partial {{\left( {{B_{n,m,k}}} \right)}^*}}}{{\partial {r_l}}} = {\left( {\frac{{\partial {B_{n,m,k}}}}{{\partial {r_l}}}} \right)^*},
\end{equation}
\begin{equation}\label{equ:B_first_order}
	\frac{{\partial {B_{n,m,k}}}}{{\partial {r_l}}} = ( { - j2\pi n\Delta f\frac{1}{c}} ){B_{n,m,k}},
\end{equation}
we can then obtain
\begin{equation}\label{equ:ln_derivative_results}
	\frac{{\partial \ln {p_{\bf{h}}}\left( {\bf{h}} \right)}}{{\partial {r_k}}} =  - \frac{1}{{\sigma _N^2}}\sum\limits_{n,m}^{} {2{\mathop{\rm Re}\nolimits} \{ {( {j2\pi n\Delta f\frac{1}{c}} )h_{n,m}^*{B_{n,m,k}}} \}}.
\end{equation}

Based on \eqref{equ:B_first_order} and \eqref{equ:ln_derivative_results}, we can obtain the second-order derivative of $\ln {p_{\bf{h}}}\left( {\bf{h}} \right)$ as
\begin{equation}\label{equ:ln_derivative2_results}
	\begin{array}{l}
		\frac{{{\partial ^2}\ln {p_{\bf{h}}}\left( {\bf{h}} \right)}}{{{\partial ^2}{r_l}}}\\
		=  - \frac{2}{{\sigma _N^2}}{\left( {2\pi n\Delta f\frac{1}{c}} \right)^2}\sum\limits_{n,m}^{} {{\mathop{\rm Re}\nolimits} \{ {{{\left| {{B_{n,m,k}}} \right|}^2} + \bar n_{t,n,m}^l{B_{n,m,k}}} \}}.
	\end{array}
\end{equation}

Since ${B_{n,m,k}} = \sqrt {P_t^U} {A_k}{e^{j2\pi m{T_s^p}{{\tilde f}_{d,k,m}}}}{e^{ - j2\pi n\Delta f\frac{{{r_k}}}{c}}}$, we can finally obtain~\eqref{equ:E_derivative2_ph}.

\end{appendices}



%

{\small
	\bibliographystyle{IEEEtran}
	\bibliography{reference}
}
\vspace{-10 mm}
\ifCLASSOPTIONcaptionsoff
  \newpage
\fi

\end{document}